\pdfoutput=1
\documentclass[]{aastex631}
\shorttitle{Intrinsic Spectral Fringe Pattern and Burst Energy Distribution of FRB 20121102A}
\shortauthors{Lyu et al.}
\graphicspath{{./}{figures/}}
\usepackage{footnote}
\usepackage{amsmath}
\usepackage{threeparttable}
\usepackage{diagbox}
\begin{document}

\title{Confining Burst Energy Function and Spectral Fringe Pattern of FRB 20121102A with MultiFrequency Observations
}
\correspondingauthor{En-Wei Liang}
\email{lew@gxu.edu.cn}
\author[0000-0002-6072-3329]{Fen Lyu}
\affiliation{Astronomical Research Center, Shanghai Science \& Technology Museum, Shanghai, 201306, China}
\affiliation{Guangxi Key Laboratory for Relativistic Astrophysics, School of Physical Science and Technology, Guangxi University, Nanning 530004, China}
\affiliation{Shanghai Astronomical Observatory, CAS, Nandan Road 80, Shanghai 200030, China}
\author[0000-0002-2585-442X]{Ji-Gui Cheng}
\affiliation{Guangxi Key Laboratory for Relativistic Astrophysics, School of Physical Science and Technology, Guangxi University, Nanning 530004, China}
\author[0000-0002-7044-733X]{En-Wei Liang*}
\affiliation{Guangxi Key Laboratory for Relativistic Astrophysics, School of Physical Science and Technology, Guangxi University, Nanning 530004, China}
\author[0000-0003-0471-365X]{Can-Min Deng}
\affiliation{Guangxi Key Laboratory for Relativistic Astrophysics, School of Physical Science and Technology, Guangxi University, Nanning 530004, China}
\author[0000-0003-4341-0029]{Tao An}
\affiliation{Shanghai Astronomical Observatory, CAS, Nandan Road 80, Shanghai 200030, China}
\author{Qing Lin}
\affiliation{Astronomical Research Center, Shanghai Science \& Technology Museum, Shanghai, 201306, China}


\begin{abstract}
The observed spectral shapes variation and tentative bimodal burst energy distribution (E-distribution) of fast radio burst (FRB) 20121102A with the FAST telescope are great puzzles. Adopting the published multifrequency data observed with the FAST and Arecibo telescopes at $L$ band and the GBT telescope at $C$ band, we investigate these puzzles through Monte Carlo simulations. The intrinsic energy function (E-function) is modeled as $dp/dE\propto E^{-\alpha_{\rm E}}$, and the spectral profile is described as a Gaussian function. A fringe pattern of its spectral peak frequency ($\nu_{\rm p}$) in 0.5-8 GHz is inferred from the $\nu_{\rm p}$ distribution of the GBT sample. We estimate the likelihood of $\alpha_{\rm E}$ and the standard deviation of the spectral profile ($\sigma_{\rm s}$) by utilizing the Kolmogorov--Smirnov (K-S) test probability for the observed and simulated specific E-distributions. Our simulations yields $\alpha_{\rm E}=1.82^{+0.10}_{-0.30}$ and $\sigma_{\rm s}=0.18^{+0.28}_{-0.06}$ ($3\sigma$ confidence level) with the FAST sample. These results suggest that a single power-law function is adequate to model the E-function of FRB 20121102A. The variations of its observed spectral indices and E-distributions with telescopes in different frequency ranges are due to both physical and observational reasons, i.e. narrow spectral width for a single burst and discrete $\nu_{p}$ fringe pattern in a broad frequency range among bursts, and the selection effects of the telescope bandpass and sensitivity. The putative $\nu_{p}$ fringe pattern cannot be explained with the current radiation physics models of FRBs. Some caveats of possible artificial effects that may introduce such a feature are discussed.
\end{abstract}




\section{Introduction} \label{sec:intro}
Fast radio bursts (FRBs) are bright (typical fluence of $\sim$Jy ms and brightness temperature of $T_{\mathrm{B}}\geq10^{35}{\rm K}$), enigmatic millisecond-duration radio bursts (\citealt{2007Sci...318..777L,2019ARA/&A..57..417C,2022A&ARv..30....2P}). More than 800 FRBs have been detected so far \citep{2016PASA...33...45P,2021ApJS..257...59C}\footnote{https://www.herta-experiment.org/frbstats/catalogue}. A high dispersion measure (88-3038 pc cm$^{-3}$; \citealt{2021ApJ...910L..18B,2021ApJS..257...59C}) of most FRBs \footnote{https://www.chime-frb.ca/catalog} indicates their extragalactic origin, which is confirmed with the host galaxy identification for some FRBs \citep{2020Natur.581..391M}. Their progenitors are thought to be flaring magnetars \citep{2014MNRAS.442L...9L,2017ApJ...843L..26B,2019MNRAS.485.4091M,2020ApJ...896..142B,2020MNRAS.498.1397L}, the giant pulses of young pulsars \citep{2010vaoa.conf..129P,2016MNRAS.462..941L,2016MNRAS.458L..19C,2021FrPhy..1624503L}, neutron stars \citep{2016ApJ...822L...7W,2016ApJ...823L..28G,2016ApJ...829...27D,2020ApJ...893L..26I}, the crust of strange stars \citep{2021Innov...200152G}, or primordial black hole mergers \citep{2018PhRvD..98l3016D,2021PhRvD.103l3030D}. The possible association between FRB 200428 and the Galactic magnetar SGR 1934 + 2154 favors the magnetar origin for at least some FRBs \citep{2020Natur.587...54C,2020Natur.587...59B}.  

The observed FRB spectrum strongly varies. Describing the spectrum with a single power-law function, $S_{\nu} \propto \nu^{\beta}$, \cite{2019ApJ...872L..19M} showed that $\beta$ varies from -15 to 10 among 23 FRBs observed with the Australian SKA Pathfinder (ASKAP). CHIME observations show that an FRB with a longer duration tends to have a narrower spectral shape, which often peaks in the instrumental bandpass due to the instrument selection effect; and a shorter burst tends to have a power-law-like spectrum covering most or all of the instrumental bandpass  (e.g.,  \citealt{2016ApJ...833..177S,2019ApJ...885L..24C,2021ApJ...923....1P,2021ApJS..257...59C}). Bright FRBs observed by ASKAP exhibit strong similar spectral modulation with unclear 
origins \citep{2018Natur.562..386S,2019ApJ...872L..19M}. It is uncertain whether the observed diverse spectra are due to their intrinsic properties or affected by the propagation effects (such as diffractive scintillation; \citealt{2016MNRAS.461.1498M,2019ApJ...872L..19M,2020ApJ...892L..10Y,2021ApJ...920...54W,2021ApJ...920L..18A}). The radiation physics of FRBs is extensively discussed (see \citealt{2020Natur.587...45Z,2021SCPMA..6449501X} for recent reviews). The proposed models \citep{2019PhR...821....1P} can be grouped into two types, i.e. synchrotron maser emission models \citep{2014MNRAS.442L...9L,2017ApJ...843L..26B,2020ApJ...896..142B,2019MNRAS.485.4091M,2020ApJ...899L..27M,2020ApJ...897....1L} and the curvature radiation models \citep{2018ApJ...868...31Y,2018ApJ...852..140W,2020MNRAS.498.1397L}.

Among the detected FRBs, most of them are one-off events, and about two dozen exhibit repeating behaviors. It remains an open question whether or not all FRBs are repeaters \citep{2018ApJ...854L..12P,2019MNRAS.484.5500C,2021ApJ...906L...5A,2022ApJ...926..206Z}. 
FRB 20121102A is the first discovered repeating FRB \citep{2016Natur.531..202S}. It has a burst rate as high as $\sim 122$  hr$^{-1}$ \citep{2021Natur.598..267L}. It even exhibits a possible periodic activity of $\sim$ 160-day  \citep{2020MNRAS.495.3551R,2021MNRAS.500..448C}. It locates in the star-forming region of a dwarf host galaxy at redshift $z \simeq 0.193$ \citep{2017ApJ...834L...7T}. The observed high rotation measure ($\mathrm{RM} \sim 10^{5}\, \mathrm{rad} \,\mathrm{m}^{-2}$) indicates that its environment is extremely magnetized \citep{2018Natur.553..182M}. Moreover, it is one of only two identified FRBs associated with a compact, luminous persistent radio source (PRS; \citealt{2017Natur.541...58C}).

FRB 20121102A is one of the most extensively monitored FRB sources with different telescopes at frequencies from 0.5 to 8 GHz with a wealth of observational features (e.g., \citealt{2014ApJ...790..101S,2016ApJ...833..177S,2016Natur.531..202S,2017Natur.541...58C,2017ApJ...834L...7T,2017ApJ...846...80S,2018ApJ...866..149Z,2018Natur.553..182M,2019ApJ...882L..18J,2019ApJ...877L..19G,2019A&A...623A..42H,2020MNRAS.495.3551R,2020ApJ...897L...4M,2020ApJ...891L...6F,2020MNRAS.496.4565C,2020A&A...635A..61O,2021Natur.598..267L,2022MNRAS.515.3577H}).  Observations indicate that the burst dynamic spectra of FRB 20121102A are highly variable. They are usually described with a Gaussian function. Note that a ``spectral index" is usually for a power-law function spectrum. We prefer adopting ``spectral slope" instead of ``spectral index" for avoiding any confusion about the Gaussian spectral model hereafter ( please refer to details under bullet point 4 in Section 3.2). Both the spectral width and peak frequency vary among bursts detected in different bandpass and different observation sessions \citep{2016Natur.531..202S,2016ApJ...833..177S,2017ApJ...850...76L,2018ApJ...863....2G,2018ApJ...866..149Z,2022MNRAS.515.3577H}. For example, the spectral indices of the bursts of FRB 20121102A observed with the Arecibo telescope at L-band ($\sim$1.4 GHz) vary from $-10.4\pm 1.1$ to $13.6\pm 0.4$ \citep{2016Natur.531..202S}. In addition, the bursts seem to be active in some specific frequencies. Search for simultaneous bursts in different frequencies have been conducted by some groups (e.g. \citealt{2016ApJ...833..177S,2017ApJ...850...76L,2019ApJ...877L..19G,2019A&A...623A..42H,2020ApJ...897L...4M,2020MNRAS.496.4565C}). \cite{2017ApJ...850...76L} presented results from a multi-telescope campaign of FRB 20121102A using the VLA at 3 GHz and 6 GHz, the Arecibo telescope at 1.4 GHz, the Effelsberg telescope at 4.85 GHz, the first station of the Long Wavelength Array (LWA1) at 70 MHz, and the Arcminute Microkelvin Imager Large Array (AMI-LA) at 15.5 GHz. Four of the nine bursts detected with the VLA had simultaneous observing coverage at different frequencies. Only one was detected simultaneously at two different observing frequencies with Arecibo (1.15-1.73 GHz) and the VLA (2.5-3.5 GHz), and none were detected during simultaneous LWA1, Effelsberg, or AMI-LA observations, despite the instantaneous sensitivities of these telescopes are better than or comparable to the VLA. \cite{2019ApJ...877L..19G} showed that among 41 bursts detected with Arecibo at 1.4 GHz, no bursts were seen with the VLA during their simultaneous observations. \cite{2015MNRAS.452.1254K} conducted a 1446-hour survey for FRBs at 145~MHz, covering a total of 4193 square degrees in the sky with the LOFAR radio telescope. No FRBs above a signal-to-noise threshold of 10 were detected. \cite{2019A&A...623A..42H} performed $\sim 20$ h of simultaneous observations with the Effelsberg 100 m radio telescope and LOFAR to search for burst activity of FRB 20121102A at 1.4 GHz and 150 MHz. They found nine bursts at 1.4 GHz but no low-frequency burst was simultaneously detected with LOFAR. Note that LOFAR detected bursts from repeating FRB 20180916B down to 0.11 GHz \citep{2021Natur.596..505P,2021ApJ...923....1P}. The non-detection of FRB 20121102A with LOFAR implies that FRB20121102A is not very active at this low frequency. 
\cite{2020ApJ...897L...4M} reported six bursts from FRB 20121102A at the 2.25 GHz frequency band, but none of these bursts is detected in the 8.36 GHz band during a campaign of 5.7 hr continuous simultaneous observation using the 70 m Deep Space Network radio telescope.  

Extensive multi-frequency observations of FRB 20121102A make it the best candidate for revealing the radiation physics of FRBs. More interestingly, a tentative bimodal distribution of the specific isotropic equivalent energy ($E_{\mu_{\rm c}}$) at the central frequency ($\mu_{\rm c}$) is found from 1652 bursts observed with the FAST telescope \citep{2021Natur.598..267L}. Note that the estimate of the burst energy depends on the burst spectrum and the bandwidth of the telescope. \cite{2021ApJ...920L..18A} argued that the bimodal feature disappears if the burst energy is calculated over the entire FAST bandpass rather than on the center frequency of the instrument. That motivates us to explore the intrinsic spectrum and energy function (E-function) underlying this phenomenon. In this paper, we employ large burst samples of FRB 20121102A observed with the FAST and Arecibo telescopes at L-band (1-2 GHz) as well as the GBT Telescope at C-band (4-8 GHz) to investigate the intrinsic spectra of the bursts from FRB 20121102A via Monte Carlo simulations over a broadband frequency range. Our paper is organized as follows. Our selected samples are presented in Section 2. Our simulations are reported in Section 3. Discussion and conclusions are given in Sections 4 and 5, respectively. Throughout, we adopt a flat $\Lambda$CDM universe with the cosmological parameters $H_{0}$=67.7$\mathrm{~km}$ $\mathrm{~s}^{-1}$ $\mathrm{Mpc}^{-1}$, $\Omega_{m}=0.31$ \citep{2016A/&A...594A..13P}.

\section{Data}\label{sec:Sample}
As mentioned above, FRB 20121102A has been extensively observed with multi-frequency telescopes in different fluence thresholds, which is estimated as \citep{2016MNRAS.458..708C,2021ApJ...923..230L},
 \begin{equation}
F_{\rm th}=\frac{T_{\text {sys }} \kappa <S/N>}{G \sqrt{B \Delta \tau N_{p}}}, \label{th}
\end{equation}
where $T_{\text {sys }}$ is the system temperature in the unit of Kelvin (K), $\kappa$ is a digitization factor, $<S/N>$ is the signal-to-noise ratio, $G$ is the system gain in the unit of $\mbox{K Jy}^{-1}$, $B$ is the bandwidth in the unit of Hz, $\Delta \tau$ is the integration time in seconds, and $N_{p}$ is the number of polarizations summed. The observed specific energy ($E_\nu$) at the peak frequency $\nu$ which is given by \cite{2018ApJ...867L..21Z},
\begin{equation}
E_{\nu}^{\rm obs}
\simeq
\left(10^{39} {\rm erg}\right)
\frac{4 \pi}{1+z}\left(\frac{D_{\rm L}}{10^{28} {\rm ~cm}}\right)^{2}
\left(\frac{F^{\rm obs}_{\nu}}{{\rm Jy} \cdot {\rm ms}}\right)\left(\frac{\nu}{\rm GHz}\right), \label{Energy}
\end{equation}
where $F^{\rm obs}_{\nu}$ is the specific fluence at $\nu$ and $D_L$ is the luminosity distance ($D_L$ = 972 Mpc for FRB 20121102A). Hereafter, we denote the simulated observable with a superscript ``sim" and denote the observational data with a superscript ``obs". We employ the data of FRB 20121102A observed with FAST, Arecibo, and GBT telescopes for our analysis \citep{2018ApJ...866..149Z,2021Natur.598..267L,2022MNRAS.515.3577H}. We describe the selected samples below.

As the largest single-dish telescope in the world, FAST has a 19-beam array receiver in L-band ranging from 1.05 GHz to 1.45 GHz with the central frequency of $\mu_{\rm c}=1.25$ GHz. Its frequency resolution is 0.122 MHz. \cite{2021Natur.598..267L} reported the results of a monitoring campaign for FRB 20121102A with the FAST telescope. During an observational period of 47 days, which took place from August 29 to October 29, 2019, the total observational time is 59.5 hours. A sample of 1652 bursts (the FAST sample) with a fluence criterion of $F^{\rm FAST}_{\rm th}=0.015$ Jy ms in the $7\sigma$ confidence level, assuming 1 ms burst duration. The frequency coverage ($\Delta\nu^{\rm obs}$) of these bursts ranges from 0.1 GHz to 0.5 GHz. The peak frequency ($\nu_{\rm p}^{\rm obs}$) of the spectrum is not available in the FAST sample. \cite{2021Natur.598..267L} calculated the specific burst energy $E^{\rm obs}_{\mu_{\rm c}}$ at $\mu_{\rm c}$. We collect the data of $E^{\rm obs}_{\mu_{\rm c}}$, $\Delta\nu^{\rm obs}$, and the burst duration ($T^{\rm obs}$) from \cite{2021Natur.598..267L}.

The Arecibo telescope is sensitive at L-band in the frequency range of 1.15-1.73 GHz in 64 channels. Its central frequency is $\mu_{c}=1.44$ GHz. Its time resolution of the sampled signal is $10.24 ~\mu$s. An observational campaign for FRB 20121102A with the Arecibo telescope was conducted from November 2015 to October 2016. The total observational time is $\sim 59$ hours. A sample of 478 bursts (the Arecibo sample) was detected \citep{2022MNRAS.515.3577H}. The bursts were selected with $F^{\rm Arecibo}_{\rm th}=0.057$ Jy ms in the $6\sigma$ confidence level assuming 1 ms burst duration. The isotropic equivalent burst energy ($E_{\Delta \nu}^{\rm obs}$) over the upper and lower frequency edges ([$\nu_{\text {high }}$,~$\nu_{\text {low}}$]) of these bursts are available in \cite{2022MNRAS.515.3577H}, where $\Delta\nu^{\rm obs}=\nu_{\text {high }}-\nu_{\text {low}}$. We collect the data of $E_{\Delta \nu}^{\rm obs}$, $T^{\rm obs}$, $\Delta\nu^{\rm obs}$, and burst edges ([$\nu_{\text {high }}$,~$\nu_{\text {low}}$]) of these bursts from \cite{2022MNRAS.515.3577H}.  

Green Bank Telescope (GBT) covers the C-band frequency range from 4 to 8 GHz \citep{2018ApJ...863....2G,2018ApJ...866..149Z}. The temporal and the frequency resolutions are 350 $\mu$s and 366 kHz, respectively. Estimating the GBT observational fluence threshold with Eq. (\ref{th}) by adopting the performance of the GBT at C-band as $T_{\rm sys}=25$ K, $\beta=2$, $N_{p}=2$, $G=2$ K~Jy$^{-1}$, $B=4$ GHz, and $<S/N>=6$, we obtain $F^{\rm GBT}_{\nu,\rm th}=0.0265$ Jy ms, which is roughly consistent with the full-band fluence limit (0.30 Jy ms) in \cite{2018ApJ...866..149Z}. A 5-hour observation campaign for FRB 20121102A with the GBT telescope on 2017 August 26 detected 93 bursts (the GBT sample).

Note that the bursts observed with GBT in the low-frequency end (4-4.5 GHz) were not analyzed by \cite{2018ApJ...863....2G} and \cite{2018ApJ...866..149Z} owing to the radio frequency interference (RFI). With a broad bandpass, the peak frequency of the spectrum ($\nu_{\rm p}^{\rm obs}$) and the radiating frequency range edges ($\nu_{\text {low}}$ and $\nu_{\text {high }}$), the specific observed fluence ( $F^{\rm obs}$) and the burst duration $T^{\rm obs}$ values of these bursts are available in \cite{2018ApJ...863....2G,2018ApJ...866..149Z}. We take these data from \cite{2018ApJ...863....2G,2018ApJ...866..149Z} and regard observed burst energy as the specific energy $E^{\rm obs}_{\nu_{\rm p}}$ at $\nu^{\rm obs}_{\rm p }$.

\begin{figure}[!htbp]
\centering
\includegraphics[width=0.45\linewidth]{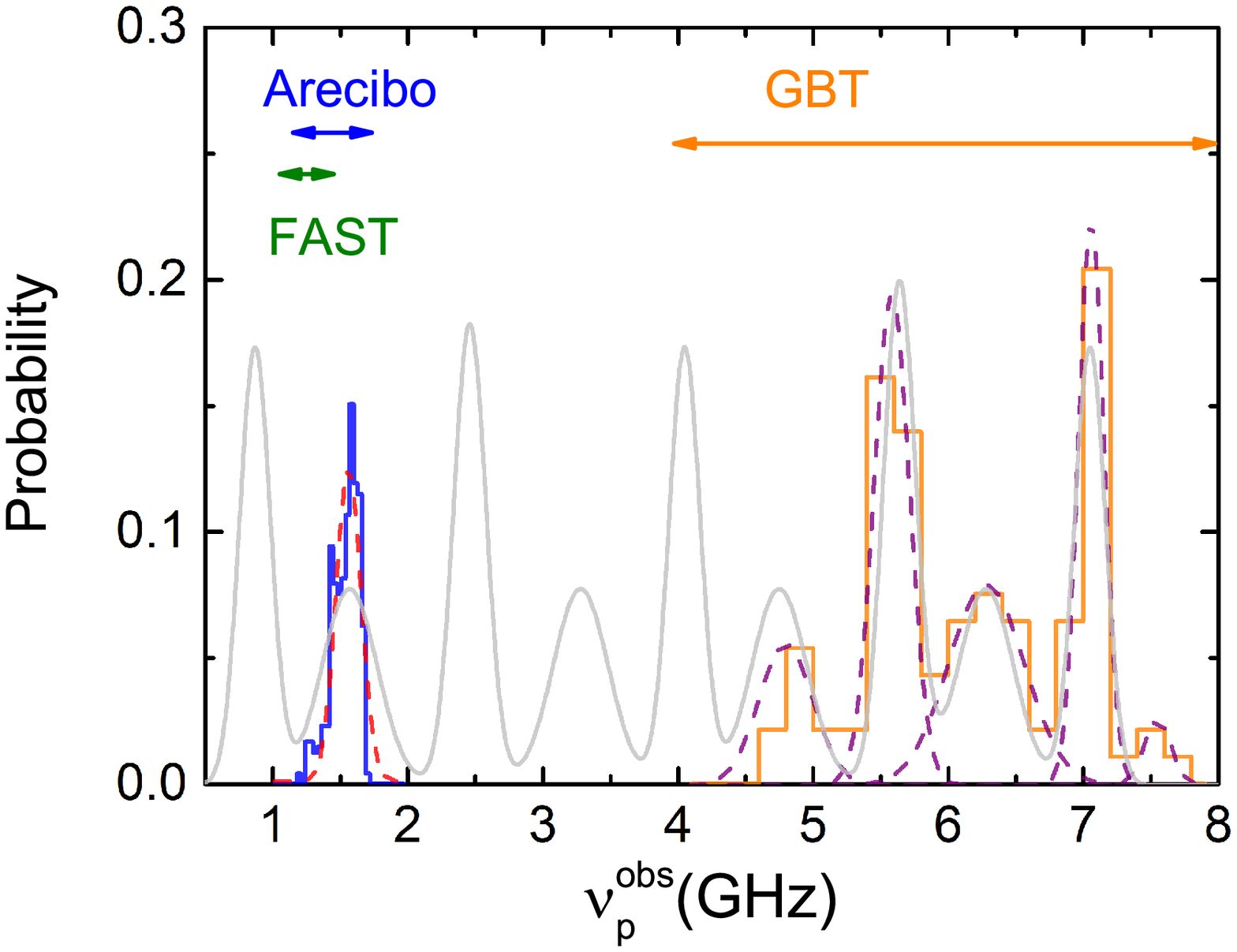}\hspace{-0.1in}
\includegraphics[width=0.45\linewidth]{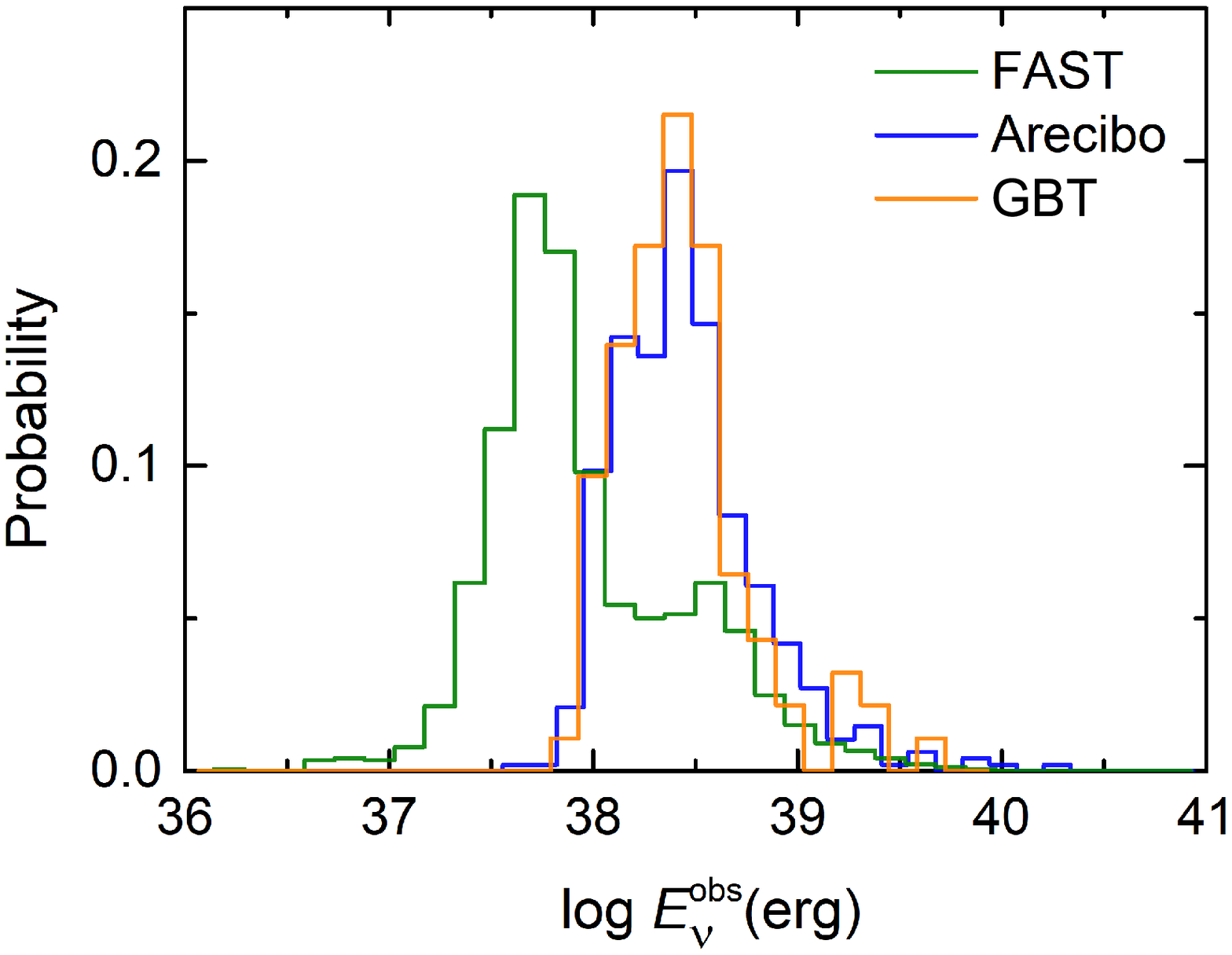}\vspace{-0.1in}
\includegraphics[width=0.45\linewidth]{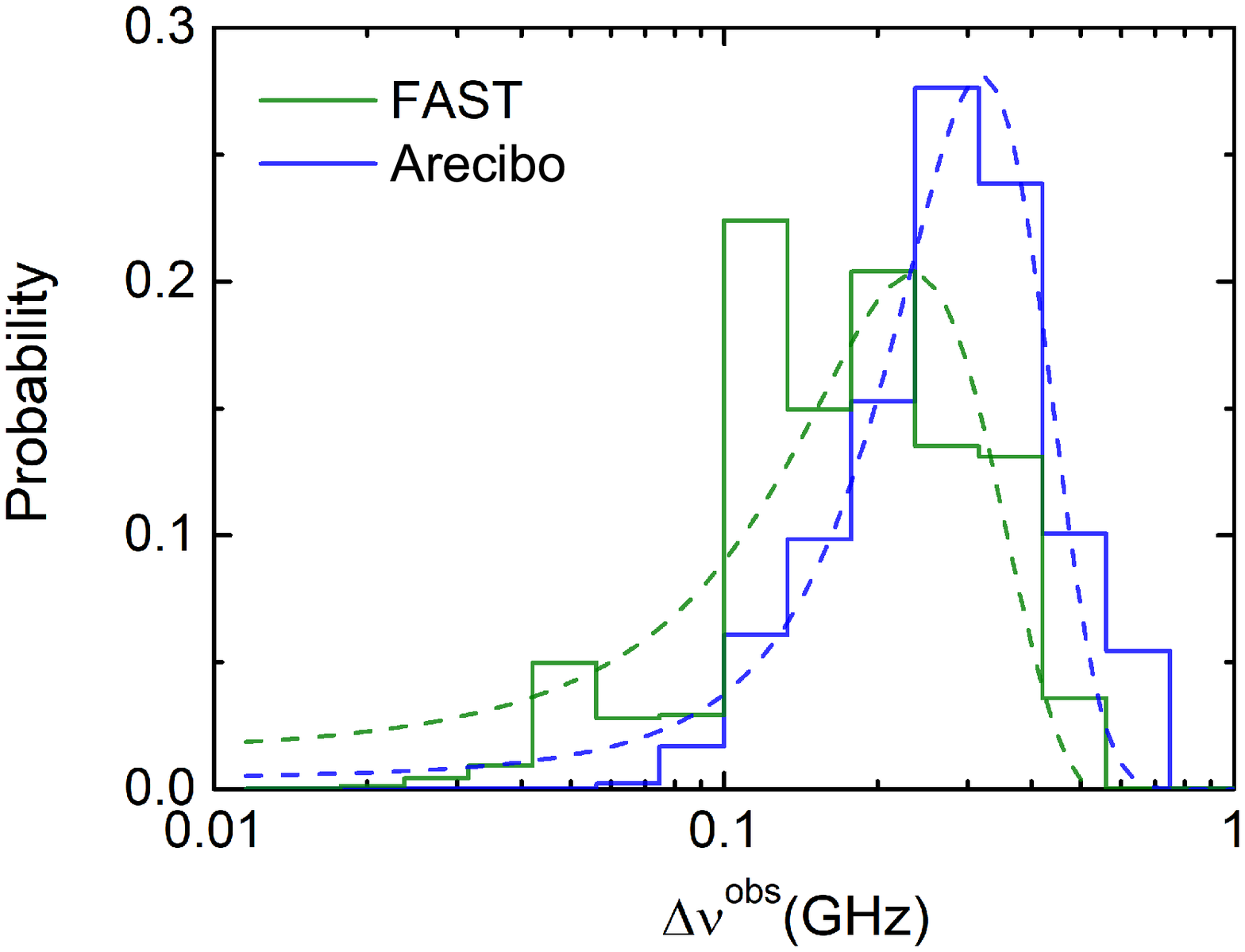}\hspace{-0.1in}
\includegraphics[width=0.45\linewidth]{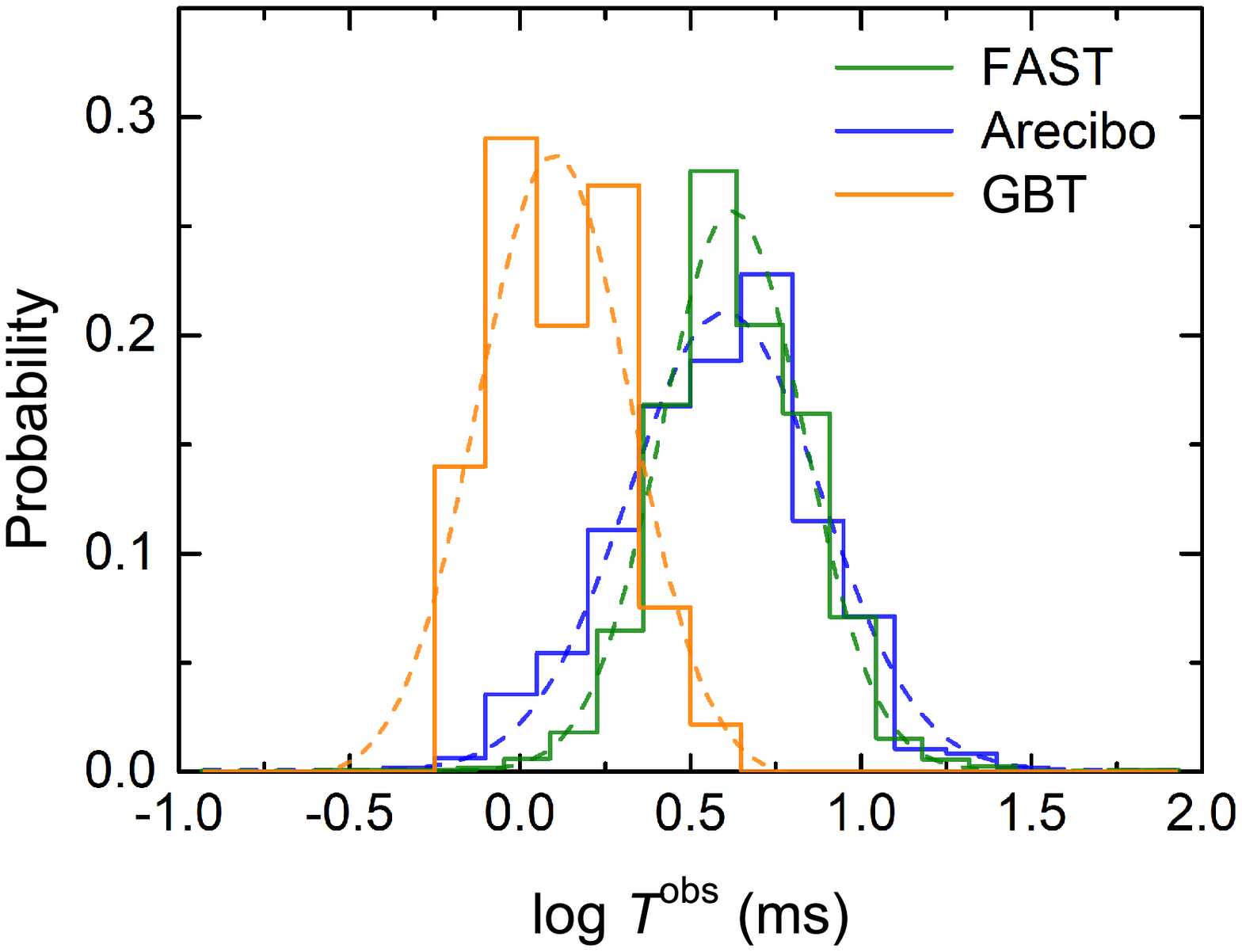}\hspace{-0.1in}
\caption{Histogram of $\nu_{\rm p}^{\rm obs}$, $\log E_{\nu}^{\rm obs}$ ($\log E_{\mu_{\rm c}}^{\rm obs}$ for the FAST sample, $\log E_{\nu_{\rm c}}^{obs}$ for the Arecibo sample, and $\log E_{\nu_{\rm p}}^{\rm obs}$ for the GBT sample), radiating frequency range $\Delta\nu^{\rm obs}$, and $\log T^{\rm obs}$. The dashed lines are our fits with a function of one or multiple Gaussian component(s). The grey line in the left-top panel is the constructed $\nu_{\rm p}$ distribution in 0.5-8 GHz inferred from the $\nu_{\rm p}^{\rm obs}$ distributions of the GBT and Arecibo samples as discussed in \S 3. The FAST, Arecibo and GBT bandpasses are also marked with arrows.  
 \label{fig:obs}}
\vspace{-0.2cm}
\end{figure}

Figure \ref{fig:obs} shows the distributions of the data for the FAST, Arecibo, and GBT samples. One can observe that the $\nu_{\text {p}}^{\rm obs}$ distribution of the GBT sample shows peaks at 4.75, 5.64, 6.28, 7.05, and 7.50 GHz. We empirically fit the distribution with multiple normal distribution functions, i.e. $dp/d_{\nu_{\rm p}}\propto \sum_i N^{\rm i}(\nu_{\rm p}; \nu_{\rm p,c,i},p_{\rm i}, \sigma_{\rm i})$, where $\nu_{\rm p,c,i}$, $p_{\rm i}$,  $\sigma_{\rm i}$ are the center value, the relative probability, and the $1\sigma$ of the component $i$th, respectively. The $\nu_{\rm p}^{\rm obs}$ distribution of the Arecibo sample can be roughly fitted with a single normal function. Our fits are also illustrated in Figure \ref{fig:obs}.

The $E^{\rm obs}_{\nu_{\rm p}}$ distribution of the GBT sample is well consistent with $E^{\rm obs}_{\nu_{\rm c}} $ distribution of the Arecibo sample. The $p_{\rm KS}$ value derived from the non-parametric Kolmogorov-Smirnov (K-S) test is 0.35. They narrowly range in $10^{38-39}$ erg. It is possible that the $\nu_{\rm p}$ of burst in the Arecibo sample would be in the Arecibo band (see further discussion in \S 3.1). Therefore, we regard  $E_{\Delta \nu}^{\rm obs}$ as $E^{\rm obs}_{\nu_{\rm p}}$ and estimate the $\nu_{\rm p}^{\rm obs}$ of these burst as $\nu^{\rm obs}_{\rm p }\simeq\nu^{\rm obs}_{\rm c }=(\nu_{\text {high }}+\nu_{\text {low}}$)/2 for the bursts in the Arecibo sample. The $E^{\rm obs}_{\mu_{\rm c}}$ distribution of the FAST sample shows a bimodal feature \citep{2021Natur.598..267L}. Its high-$E^{\rm obs}_{\mu_{\rm c}}$ also ranges in $10^{38-39}$ erg, being consistent with the $E^{\rm obs}_{\nu_{\rm p}}$ distributions of the GBT and the Arecibo samples. 

The distributions of the spectral widths of the bursts in the FAST and Arecibo samples are $\Delta\nu^{\rm obs}=0.1\sim 0.5$ GHz with a typical value of $\Delta\nu=0.23$ GHz and $\Delta\nu^{\rm obs}=0.1\sim 0.7$ GHz with a typical value of 0.32 GHz, respectively. The burst durations $T^{\rm obs}$ of the FAST and Arecibo samples are comparable, and they are much longer than that of the bursts in the GBT sample. Fitting the distribution of $\log T^{\rm obs}$ with a normal function, we have $\log (T^{\rm obs}/{\rm ms})=0.62\pm 0.25$, $0.60\pm 0.33$, and $0.10\pm 0.26$ for the FAST, Arecibo, and GBT samples, respectively. This indicates that bursts in the high frequency band are usually shorter than that in the low frequency band (e.g. \citealt{2018ApJ...863....2G}). 

\section{Simulation Analysis}
\label{sec:sim}
\subsection{Model Assumptions}
\begin{itemize}
    \item Intrinsic Energy Function (E-function): We model the intrinsic E-function of the bursts in the range of $[10^{37}, 10^{42}]$ erg with a single power-law function,
        \begin{equation}
        \frac{dp(E)}{dE}\propto E^{-\alpha_{\rm E}} \label{PL}.
        \end{equation}
    \item Intrinsic Spectral Profile: The spectra of bursts from FRB 20121102A seem to have a Gaussian envelope \citep{2017ApJ...850...76L,2018ApJ...863....2G,2019ApJ...882L..18J}. Therefore, we model the spectral profile as a Gaussian function
\begin{equation}
F_\nu
=\frac{F}{\sigma_{\rm s} \sqrt{2 \pi}} \exp \left[\frac{-\left(\nu-\nu_{\rm p}\right)^{2}}{2\sigma_{\rm s}^{2}}\right], \label{Gauss}
\end{equation}
where $\nu_{\rm p}$ is the peak frequency, $\sigma_{\rm s}$ is the standard deviation, and $F$ ($=E/4\pi D_L^2$) is the fluence of the spectrum.

\item  Intrinsic Spectral Fringe Pattern:
As shown in Figure \ref{fig:obs}, the detection probability distribution of $\nu^{\rm obs}_{\rm p }$ in the 4-8 GHz observed with the GBT telescope is not uniform. It shows a fringe feature. Each fringe component $i$ can be fitted with a normal distribution function $N_{\rm i}(\nu_{\rm p}; \nu_{\rm p,c,i},p_{\rm i}, \sigma_{\rm i})$. Since the peaks at 5.64, 6.28, and 7.05 GHz are less suffered the frequency-cut effect, we regard it as a basic block of the fringes and extrapolate it to the range from 0.5 GHz to 5 GHz. The constructed probability distribution of $\nu_{\rm p}$ in the range from 0.5 GHz to 8 GHz is also shown in Figure \ref{fig:obs}. It reads as
\begin{equation}\label{nu_p}
    dp/d{\nu_{\rm p}}\propto \sum_i N(\nu_{\rm p}; \nu_{\rm p,c,i},p_{\rm i}, \sigma_{\rm i}).
\end{equation}
The parameters of $\nu_{p,c}$, $p$, $\sigma$ of each component are summarized in Table \ref{table_global_nu_p}. Interestingly, the bandpass of the Arecibo telescope covers the $\nu_{\rm p}$ distribution peak at 1.57 GHz\footnote{Note that the relative probability of the $\nu_{\rm p}$ distribution between the GBT and Arecibo samples cannot be directly compared.} since the two telescopes have different sensitivity, indicating that the $\nu_{\rm p}$ for a large fraction of bursts in the Arecibo sample should be in the bandpass of the telescope. No burst in the 2-4.5 GHz is available in the selected FAST, Arecibo, and GBT samples. 
\end{itemize}

\begin{table}[!htbp]
    \centering
\caption{The intrinsic $\nu_{\rm p}$ distribution of FRB 20121102A in the frequency coverage [0.5, 8] GHz inferred from the observed $\nu_{\rm p}^{\rm obs}$ distributions of the GBT and Arecibo samples.}\label{table_global_nu_p}
\begin{tabular}{|c|c|c|c|}
\hline
$i$th fringe & ${p_{\rm i}}$ & $\nu_{\rm p,c,i}$&$\sigma_{\rm i}$\\
\hline
1& 0.05 &0.87 & 0.23 \\
2 &0.04 & 1.57&0.41\\
3&  0.06 & 2.46&0.24\\
4 & 0.04 & 3.28&0.41\\
5 & 0.05 & 4.05&0.23\\
6 & 0.04 & 4.75&0.41\\
7 &0.05 & 5.64&0.24\\
8 &0.04 & 6.28& 0.50\\
9 & 0.06 & 7.05&0.23\\
\hline
\end{tabular}\\
\end{table}

\subsection{Simulation Procedure}
Utilizing the FAST, Arecibo, and GBT samples, we constrain $\alpha_{\rm E}$ and $\sigma_{\rm s}$ of FRB 20121102A with Monte Carlo simulations. We outline our simulation procedure as follows.

\begin{enumerate}
    \item We assume that the $\alpha_{\rm E}$ value uniformly distributes in the range of [1,4], then randomly pick up a $\alpha_{\rm E}$ value.
    \item For a given $\alpha_{\rm E}$ value, we simulate a burst. Its energy $E^{\rm sim}$ is generated from the intrinsic $E-$function of Eq. (\ref{PL}). We assume that the simulated burst duration distributions for the FAST, Arecibo and GBT samples are the same as the observed ones (as shown in Figure \ref{fig:obs}), and generate the burst duration ($T^{\rm sim}$) from the $\log T^{\rm obs}$ probability distributions.
    \item We simulate an intrinsic radiation spectrum, which is described as Eq. (\ref{Gauss}), for a given burst. \cite{2021RNAAS...5...17F} divided the entire 4 GHz bandwidth of the GBT observation \citep{2018ApJ...863....2G} into 8 sub-bands. Each band spans $\sim$ 500 MHz. They show that the spectral width ranges from 120 to 650 MHz. Therefore, we assume that $\sigma_{\rm s}$ of the Gaussian spectrum uniformly distributes in the range of 0.1-0.8 GHz, then randomly pick up a $\sigma_{\rm s}$ value. The $\nu_{\rm p}$ value of the spectrum is generated based on Eqs. (\ref{nu_p}). We define the observable spectral range as $[\nu_1,\nu_2]$, where $\nu_1=\nu_{\rm p}-\sigma_{\rm s}$ and $\nu_2=\nu_{\rm p}+\sigma_{\rm s}$. We exclude those bursts that have $\nu_1\leq 0$. Since the bandpasses of the FAST and Arecibo telescopes are below 3 GHz and $\sigma_{\rm s}$ is limited as $<0.8$ GHz, we take only the fringe components 1-3 (0.5-3 GHz) of the $dp/d\nu_{\rm p}$ distribution for generating the $\nu_{\rm p}$ for the simulated FAST and Arecibo samples. Similarly, we take only the fringe components 5-8 (4-8 GHz) for generating the $\nu_{\rm p}$ in the simulated GBT sample.

    \item We calculate the specific energy and spectral slope of the simulated burst in the observable frequency range. We define the observable frequency range of a simulated burst in the spectral range $[\nu_1,\nu_2]$ with an instrument in the bandpass $[\mu_1,\mu_2]$ as $[\lambda_1,\lambda_2]$, where $\lambda_1=\max(\nu_1, \mu_1)$ and $\lambda_2=\min(\nu_2, \mu_2)$. We exclude those bursts whose observable ranges are out of the instrumental bandpass, i.e. $\nu_1\geq \mu_2$ or $\nu_2\leq \mu_1$. We do not discriminate the $\nu_{\rm p}$-in-band and $\nu_{\rm p}$-out-band scenarios, and calculate the specific energy at the central frequency, i.e. $\nu^{\rm sim}_c=(\lambda_1+\lambda_2)/2$, with Eq. (\ref{Energy}). We measure the global feature of an observable spectrum by calculating its spectral slope with $\beta^{\rm sim}=\log_{10}(F^{\rm sim}_{\lambda_2}/F^{\rm sim}_{\lambda_1})/\log_{10}(\lambda_2/\lambda_1)$. It presents the global shape of a spectrum, i.e. a rising, decaying, or flattening spectrum in the range $[\lambda_1,\lambda_2]$. 

    \item We check the detectability of the simulated burst with the FAST, Arecibo, and GBT telescopes. We calculate the equivalent flux density of the burst in the unit of Jy ms with $S_\nu^{\rm sim}=F^{\rm sim}/(\omega^{\rm sim}_{\rm eq} T^{\rm sim})$, where $F^{\rm sim}$ is the observable fluence given by $F^{\rm sim}=\int_{\lambda_1}^{\lambda_2} F_\nu d\nu$, and $\omega^{\rm sim}_{\rm eq}$ is the equivalent spectral width of the burst, which is defined as the full-width-half-maximum (FWHM) in the range of [$\lambda_1$,$\lambda_2$]. Note that the baseline for calculating the FWHM is taken as the minimum flux between $F(\lambda_1)$ and $F(\lambda_2)$, but not zero. A burst is detectable if its $S_\nu^{\rm sim}$ is larger than the telescope thresholds, i.e. $S^{\rm FAST}_{\nu,\rm th}=0.015$ Jy ms, $S^{\rm Arecibo}_{\nu,\rm th}=0.057$ Jy ms, and $S^{\rm GBT}_{\nu,\rm th}=0.0265$ Jy ms.

\end{enumerate}


\subsection{Results}
\begin{figure}[!htbp]
\centering
\includegraphics[width=0.3\linewidth]{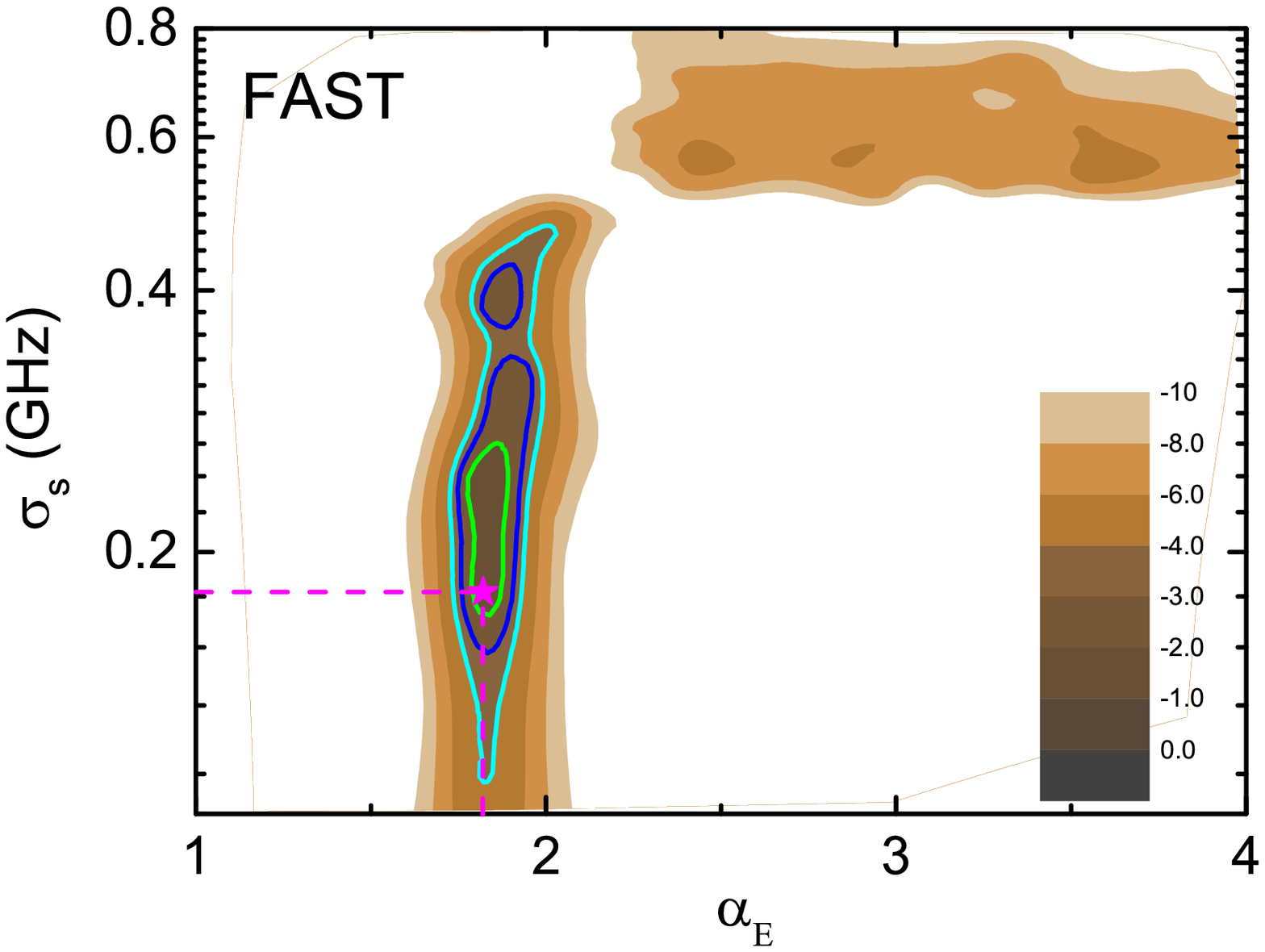}
\includegraphics[width=0.3\linewidth]{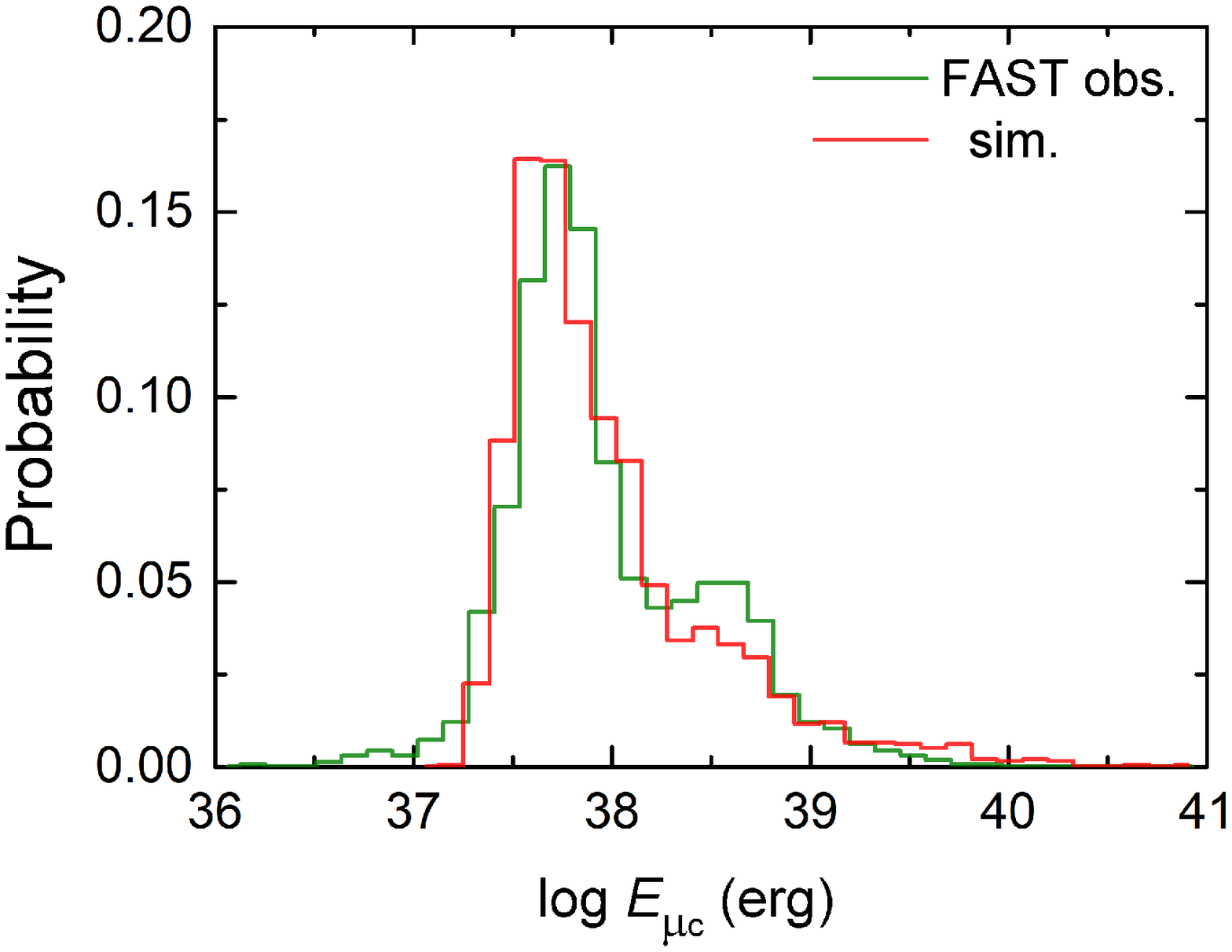}\includegraphics[width=0.3\linewidth]{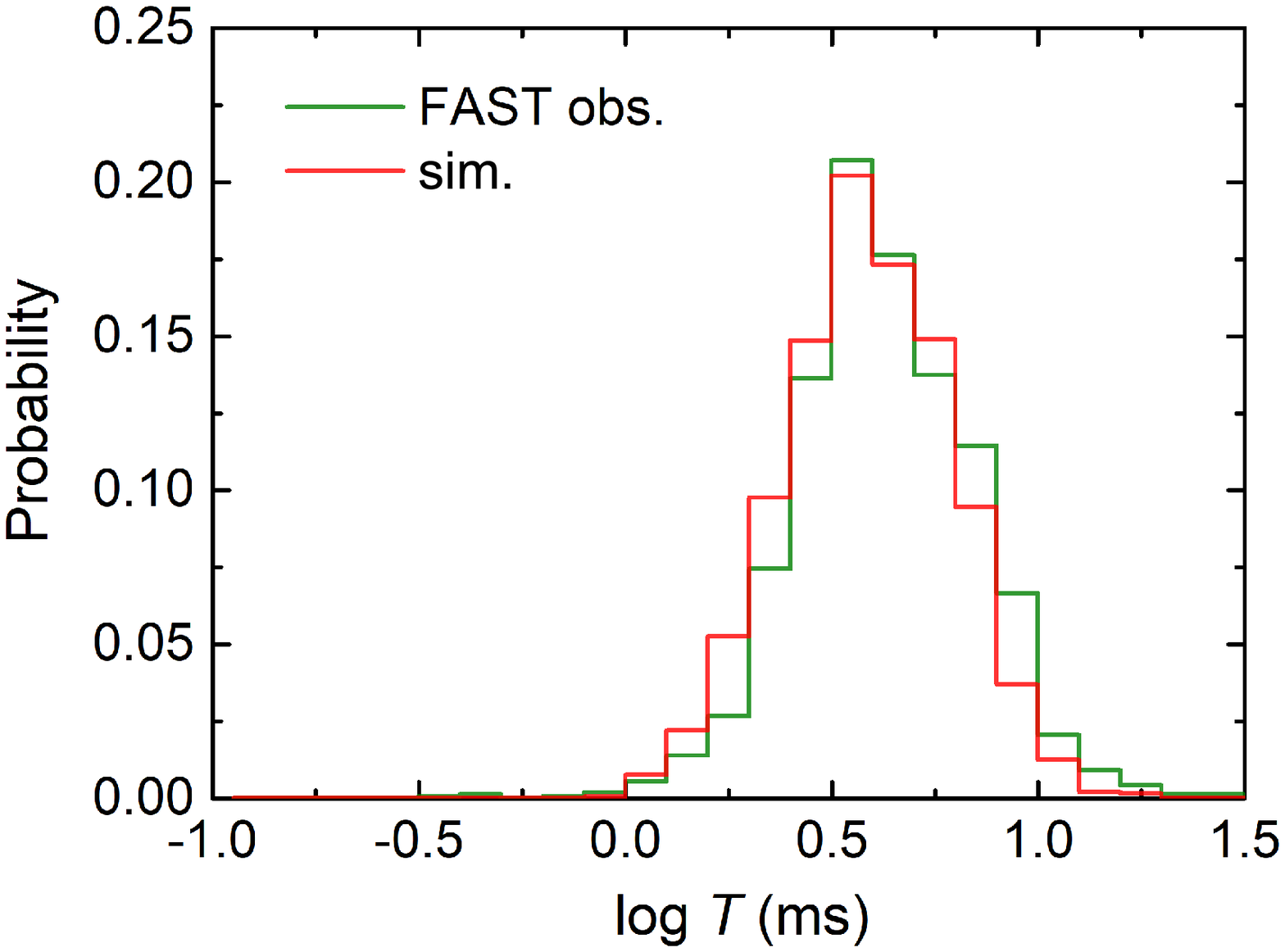}
\includegraphics[width=0.3\linewidth]{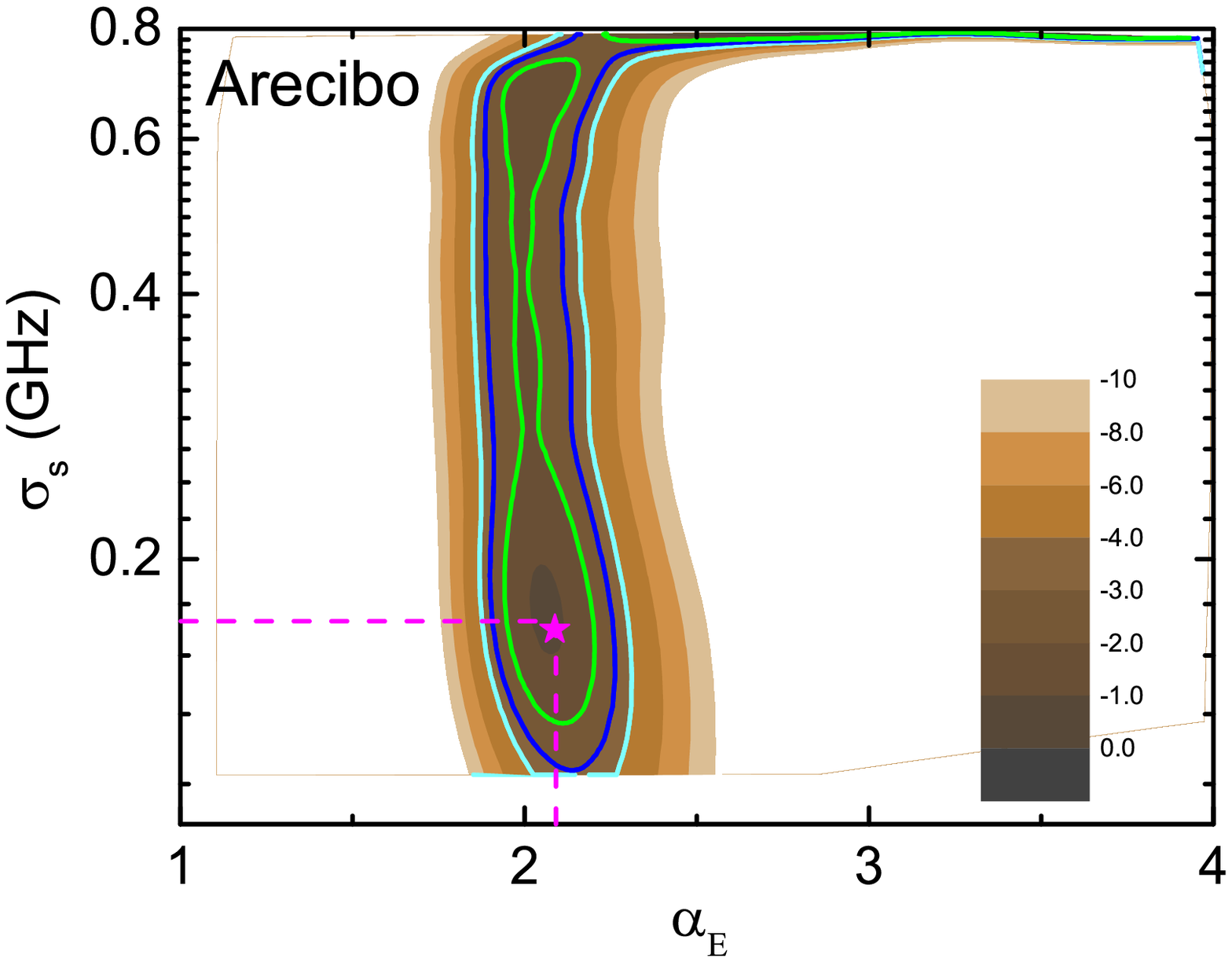}
\includegraphics[width=0.3\linewidth]{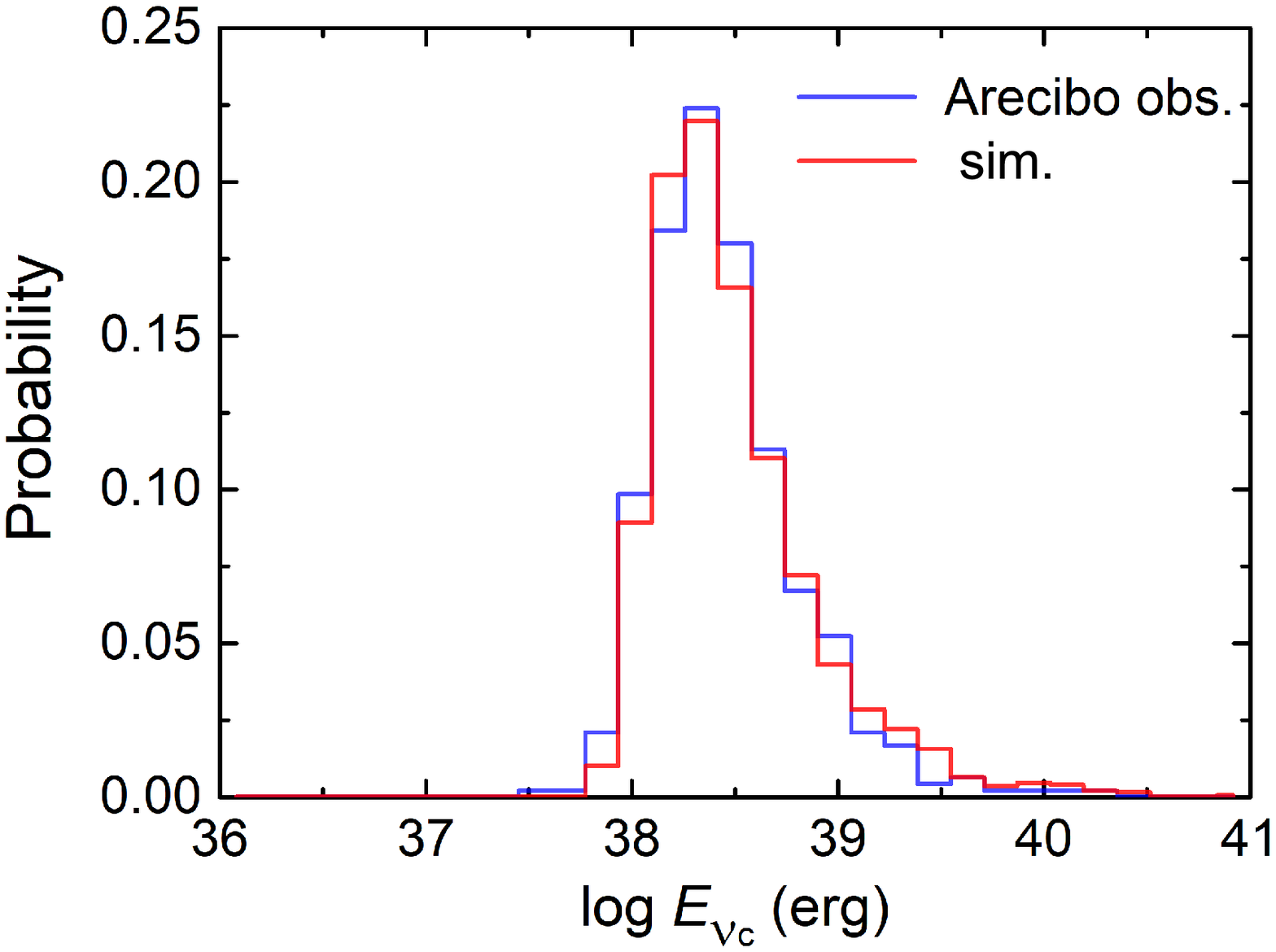}
\includegraphics[width=0.3\linewidth]{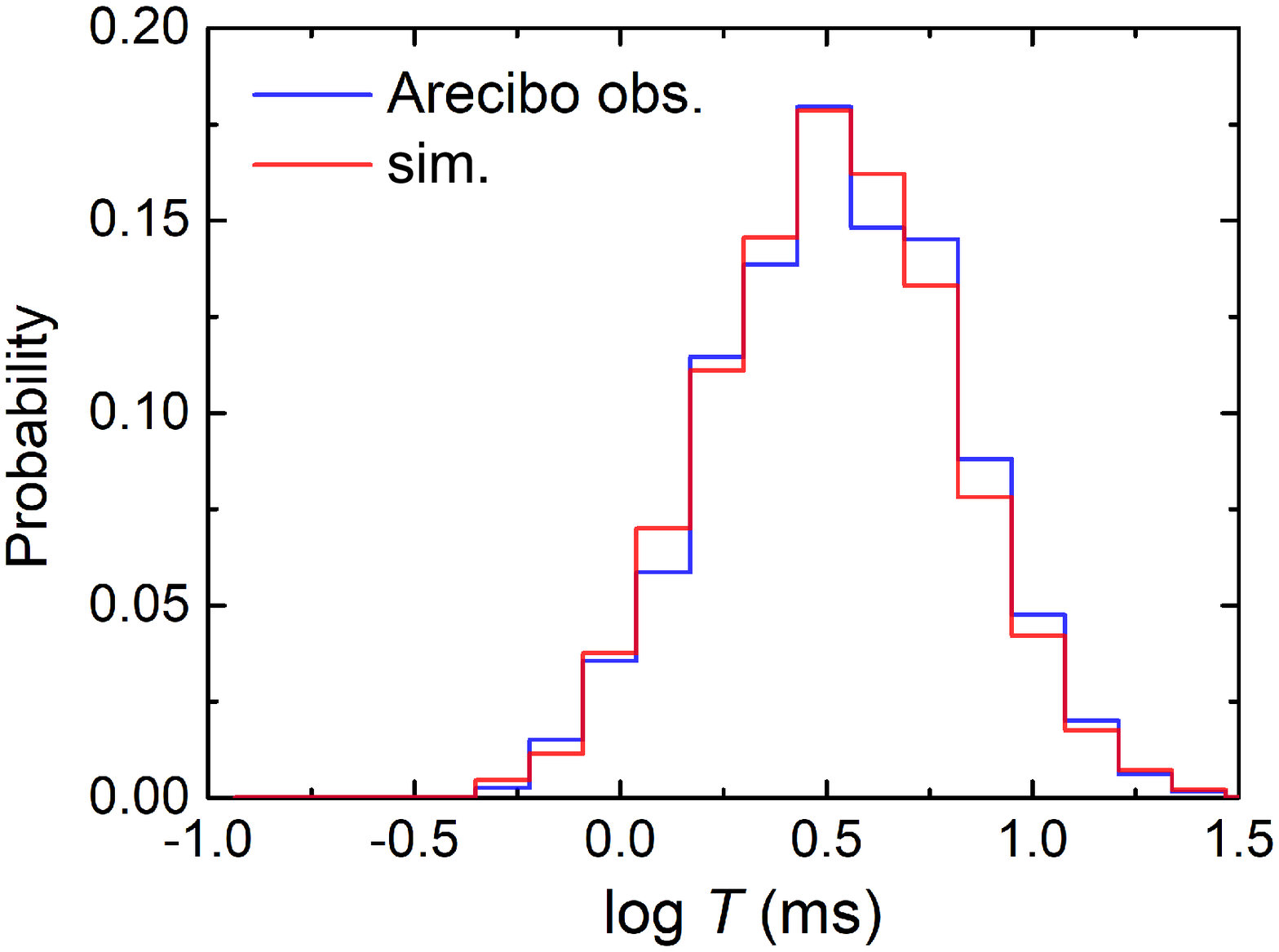}
\includegraphics[width=0.3\linewidth]{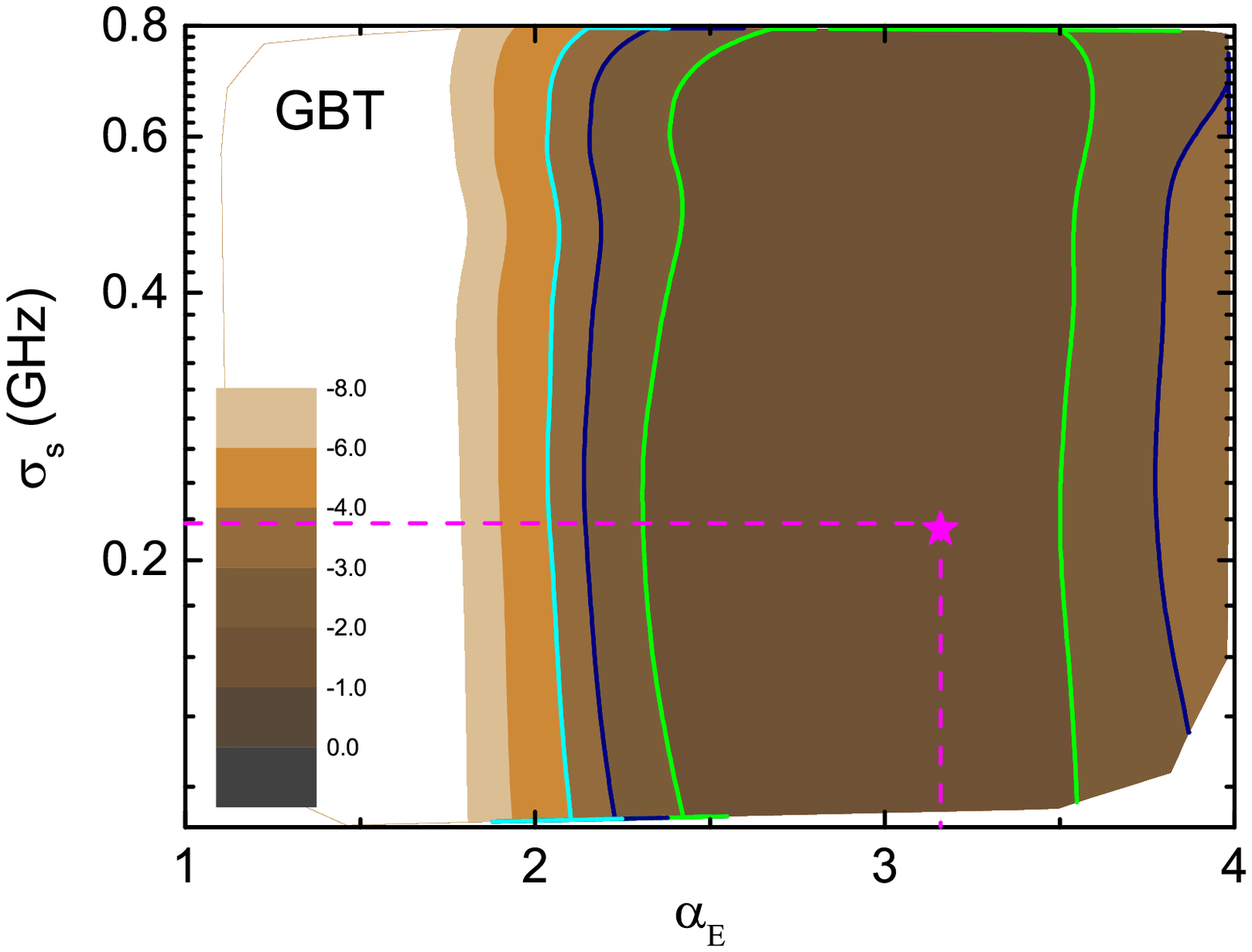}
\includegraphics[width=0.3\linewidth]{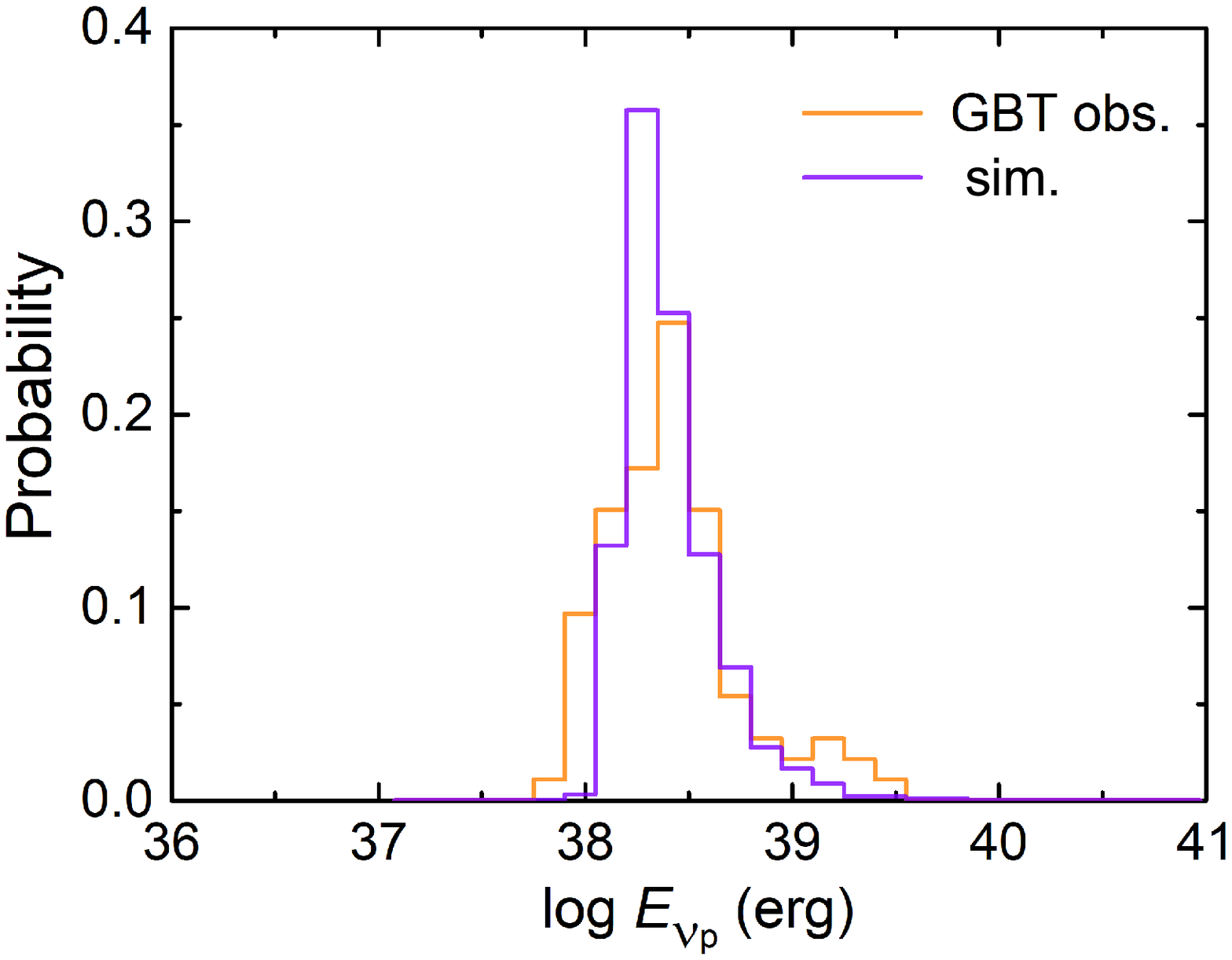}
\includegraphics[width=0.3\linewidth]{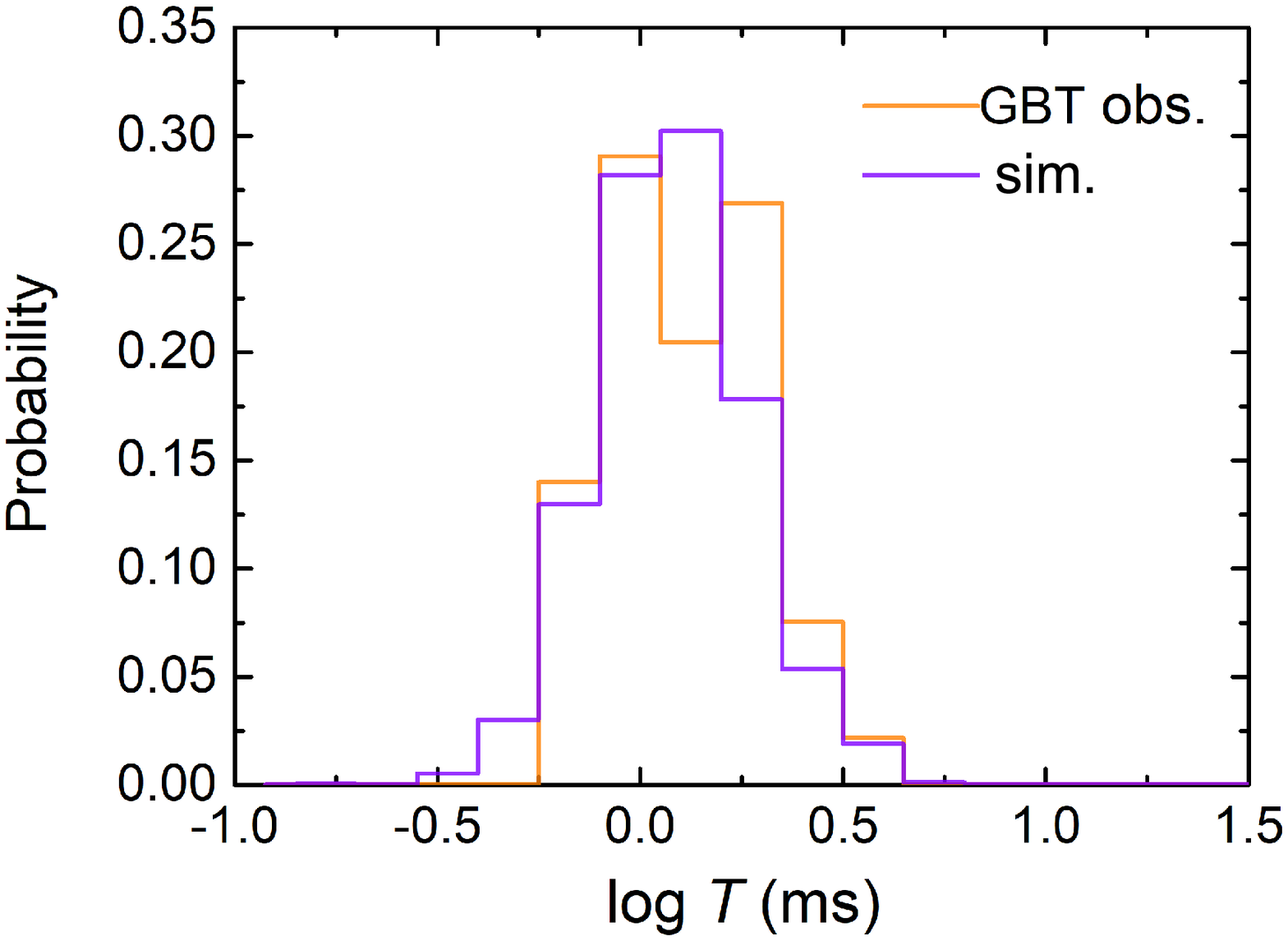}
\caption{The $\log p_{\mathrm{KS}}$ contours in the $\sigma_{\rm s}-\alpha_{\rm E}$ plane ({\em left panels}) and comparisons of the distributions of the burst specific energy (at $\mu_{\rm c}$ for the FAST sample, $\nu_{\rm c}$ for the Arecibo sample, or $\nu_{\rm p}$ for the GBT sample) and burst duration between the observed and simulated samples by adopting the maximum likelihood parameter sets of $\{\alpha_{\rm E},\sigma_{\rm s}\}$ (magenta stars in the left panels), i.e. $\{1.82,0.18~{\rm GHz}\}$ for the FAST sample ($ P_{\mathrm{KS}}$=0.08), $\{2.09,0.17~{\rm GHz}\}$ for the Arecibo sample ($ P_{\mathrm{KS}}$=0.31), and $\{3.16,0.22~{\rm GHz}\}$ for the GBT sample ($ P_{\mathrm{KS}}$=0.10). The cyan, blue, and green lines in the left panels mark the contours of $\log P_{\mathrm{KS}}$=-4, -3, and -2, respectively.
\label{fig:contours}}
\end{figure}

 Figure \ref{fig:contours} shows the contours of $\log p_{\mathrm{KS}}$ in the $\alpha_{\rm E}$-$\sigma_{\rm s}$ plane for the three samples. It is found that the observed E-distributions of the three samples can be reproduced by our simulations at a confidence level of $3\sigma$. As marked with magenta stars in Figure \ref{fig:contours}, the derived $\{\alpha_{\rm E},\sigma_{\rm s}\}$ set with the maximum likelihood is $\{1.82,0.18\}$ from the FAST sample ($p_{\rm KS}=0.08$), $\{2.09,0.17\}$ from the Arecibo sample ($p_{\rm KS}=0.31$), and $\{3.16,0.19\}$ from the GBT sample ($p_{\rm KS}=0.08$). Comparisons of the distributions of the burst specific energy (at $\mu_{\rm c}$ for the FAST sample, $\nu_{\rm c}$ for the Arecibo sample, and $\nu_{\rm p}$ for the GBT sample) and burst duration between the observed and simulated samples by adopting the derived maximum likelihood parameter sets are also shown in Figure \ref{fig:contours}. It is found that the $E_{\nu_{\rm c}}^{\rm obs}$ and $T^{\rm obs}$ distributions of the three samples are well reproduced. Note that we do not make any dependence of the burst energy $E$ (or fluence $F$) on the burst duration, and the $T^{\rm sim}$ values are bootstrapped from the  $T^{\rm obs}$ distributions. This eventually makes the distributions of $T^{\rm sim}$ and $T^{\rm obs}$ are roughly consistent, yielding $P_{\rm KS}=4.56\times 10^{-6}$, 0.23, and $3.40\times 10^{-6}$ for the FAST, Arecibo, and GBT samples, respectively.         

The $\log p_{\mathrm{KS}}$ contour for the FAST sample illustrates two distinct regions, i.e. $\{\sigma_{\rm s}< 0.5,\alpha_{\rm E} < 2.3\}$ and $\{\sigma_{\rm s}> 0.5,\alpha_{\rm E} > 2.3\}$. In the $\{\sigma_{\rm s}> 0.5,\alpha_{\rm E} > 2.3\}$ region, the $p_{\rm KS}$ values are $<10^{-4}$. In addition, as mentioned above, the spectral width of the bursts observed with GBT is typically in the range of 120 $\sim$ 650 MHz \citep{2021RNAAS...5...17F}. Therefore, the potential parameter region of $\{\sigma_{\rm s}> 0.5,\alpha_{\rm E} > 2.3\}$ is ruled out in our analysis. The most preferable parameter region is at the region of $\{\sigma_{\rm s}< 0.5,\alpha_{\rm E} < 2.3\}$, which gives $\alpha_{\rm E}=1.8^{+0.1}_{-0.3}$ and $\sigma_{\rm s}=0.18^{+0.28}_{-0.06}$ by adopting $p_{\rm KS}>10^{-4}$. The bimodal $\log E^{\rm obs}_{\mu_{\rm c}}$ distribution of the FAST sample can be reproduced with this optimal parameter set. A similar feature is also seen in the Arecibo sample, but the two-parameter regions cannot be well separated. The $\alpha_{\rm E}$ value is constrained as $\alpha_{\rm E}=2.09^{+0.30}_{-0.40}$, but $\sigma_{\rm s}$ cannot be constrained with $p_{\rm KS}>10^{-4}$. Since the size of the GBT sample is very small (93 bursts), it gives $\alpha_{\rm E}>2.1$ but loses any constraint on $\sigma_{\rm s}$ with $p_{\rm KS}>10^{-4}$. 

\begin{figure}[!htbp]
\centering
\includegraphics[width=0.35\linewidth]{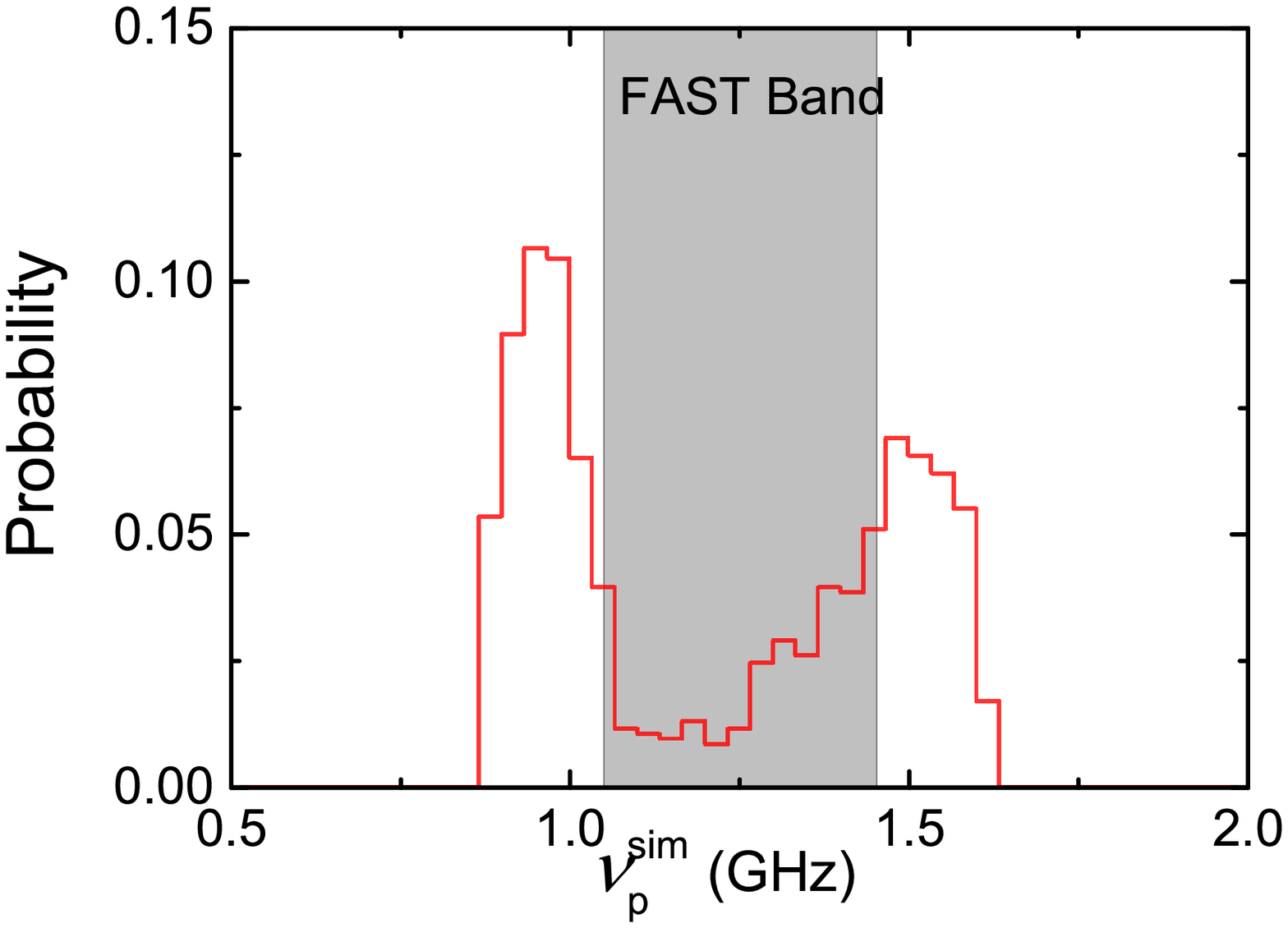}
\includegraphics[width=0.35\linewidth]{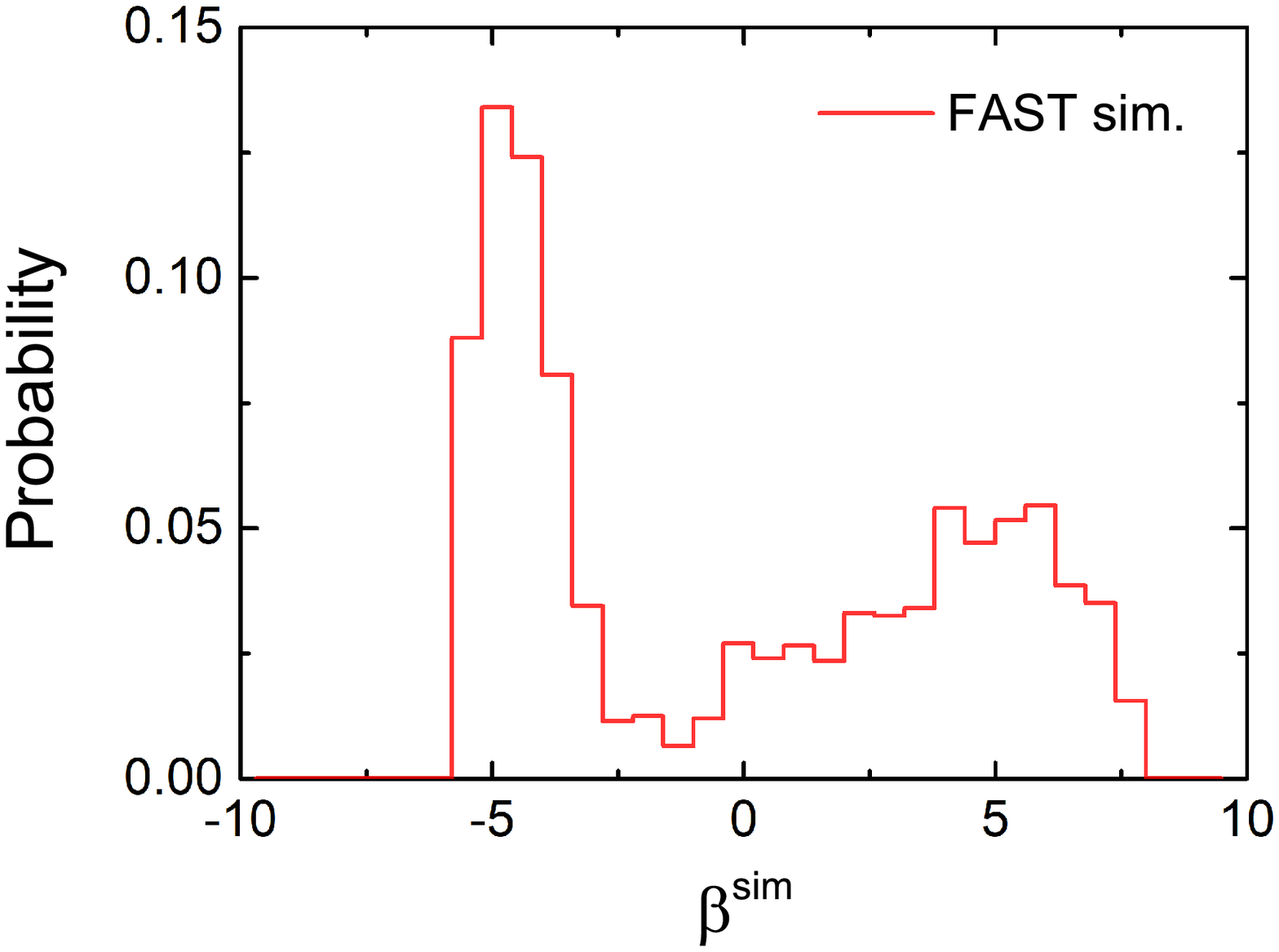}
\includegraphics[width=0.35\linewidth]{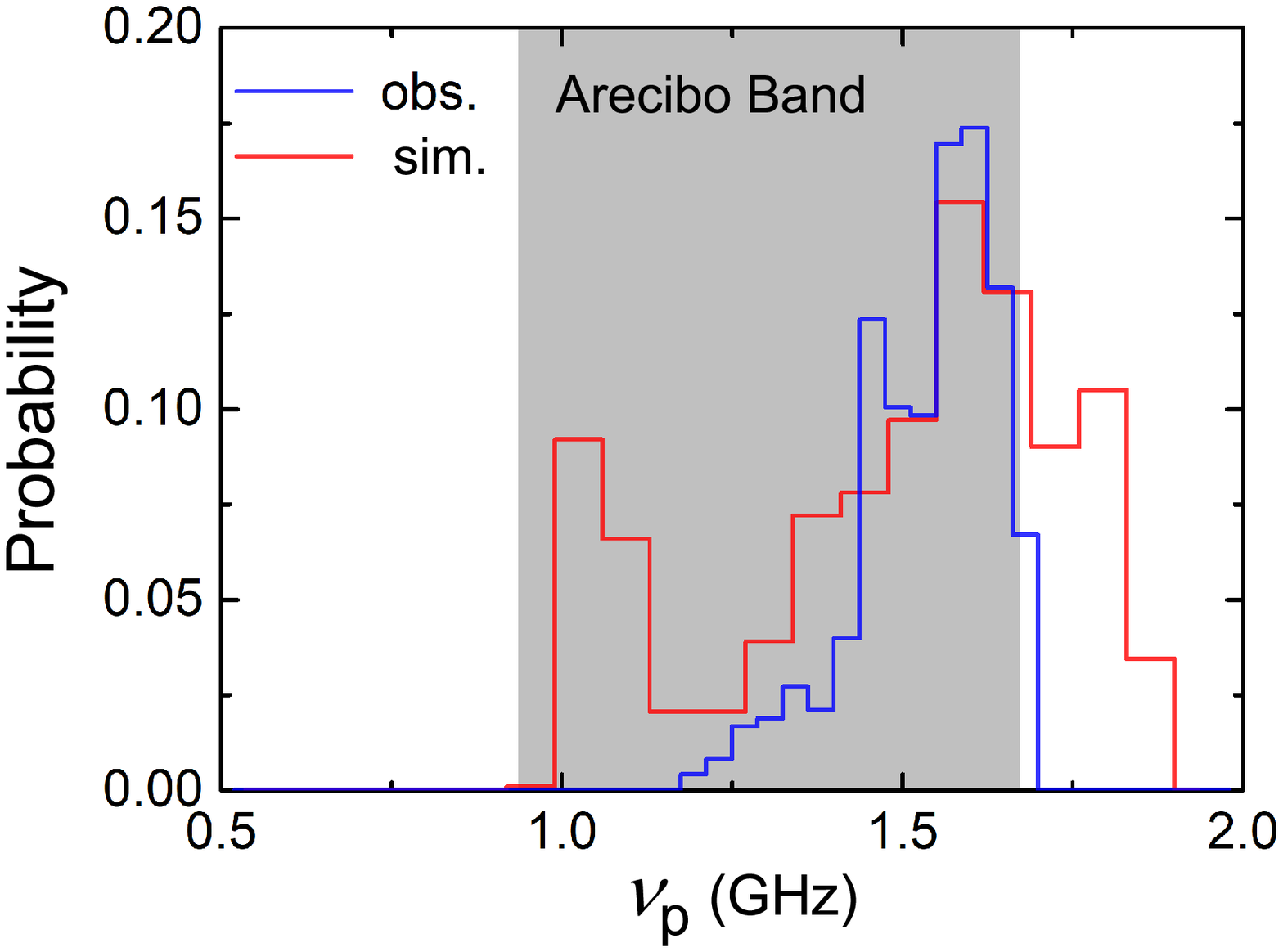}
\includegraphics[width=0.35\linewidth]{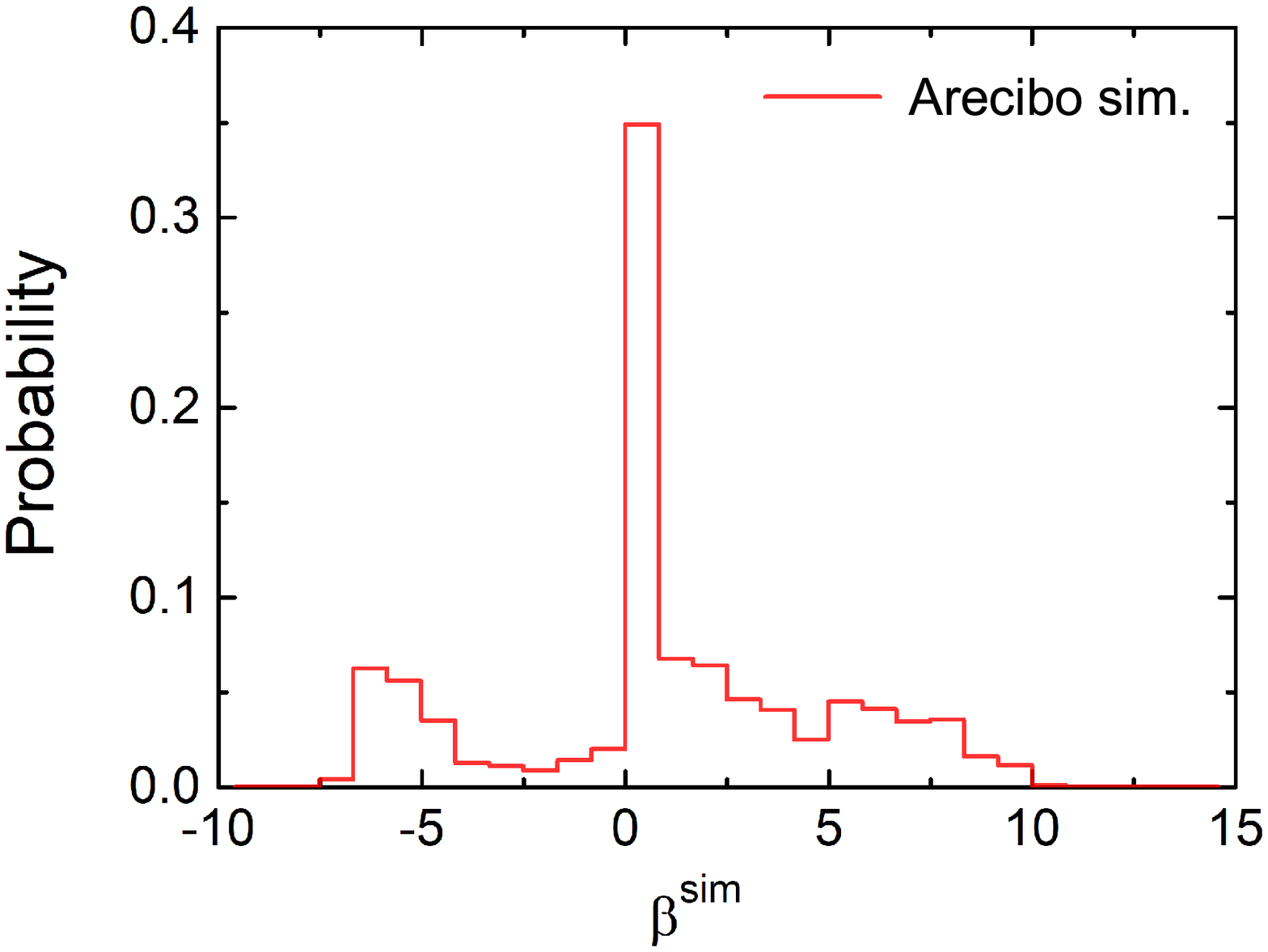}
\includegraphics[width=0.35\linewidth]{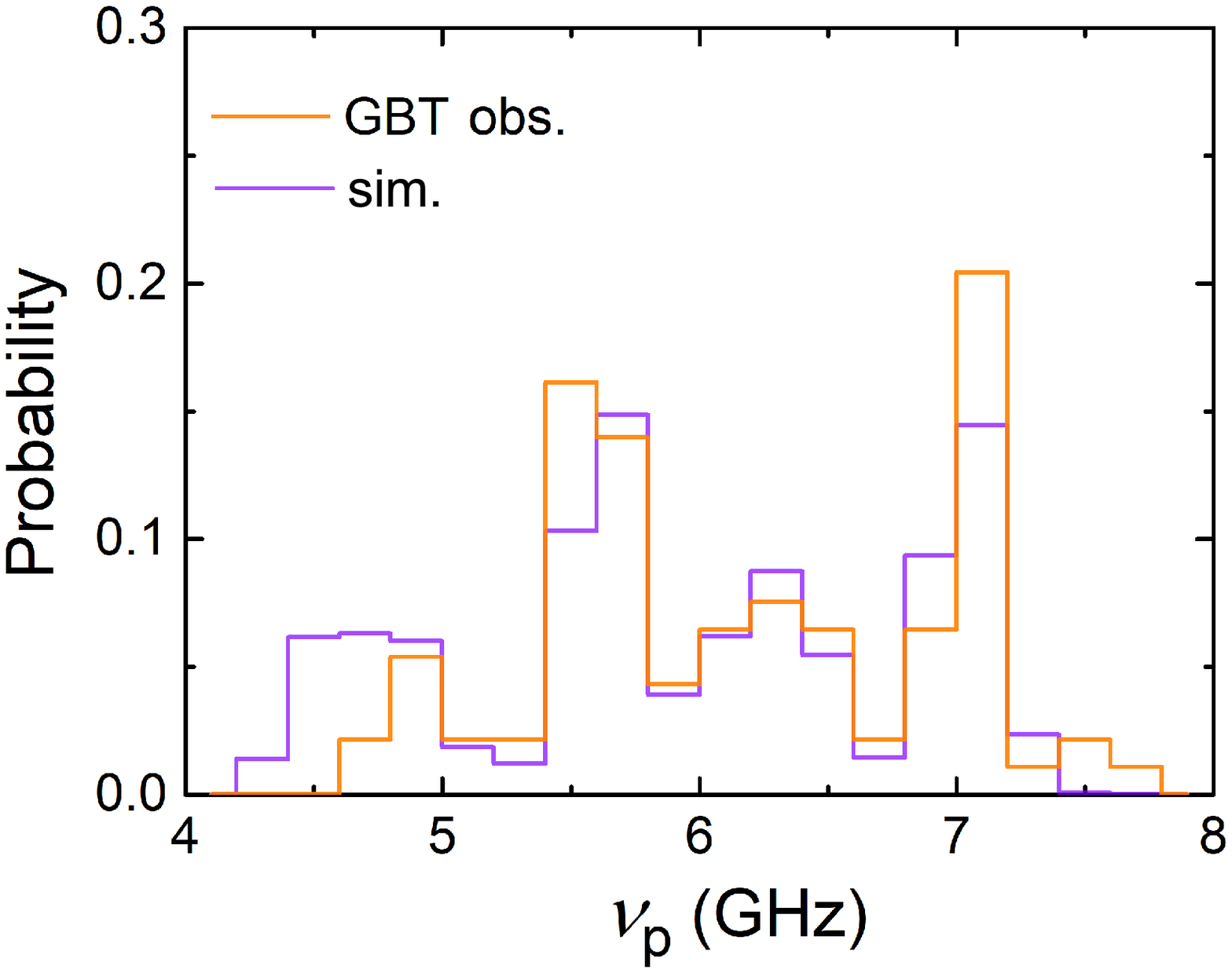}
\includegraphics[width=0.35\linewidth]{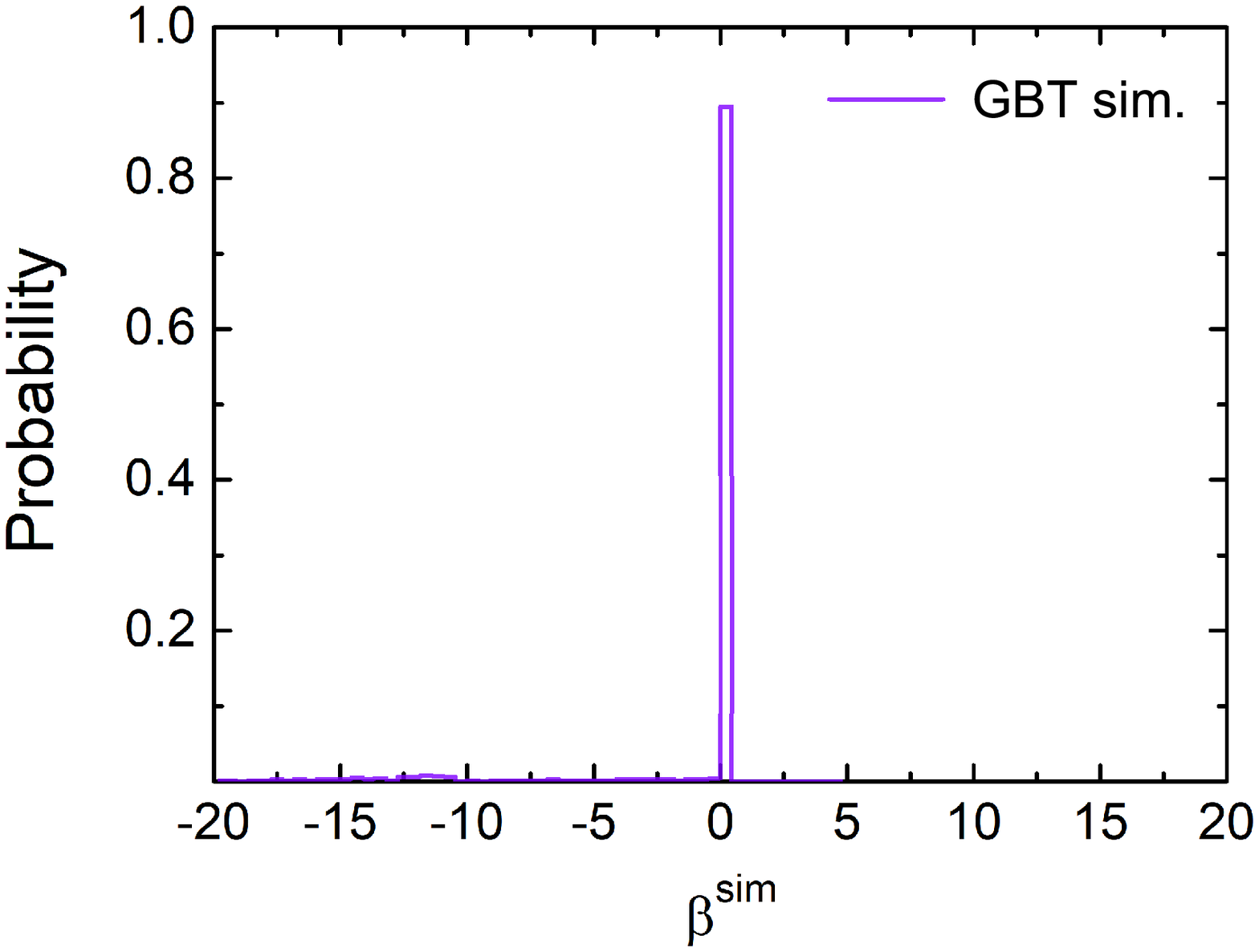}
\caption{Distributions of $\nu^{\rm sim}_{\rm p}$ and $\beta^{\rm sim}$ of the simulated samples in comparison with the observed ones if available. The simulated samples are generated by adopting the maximum likelihood parameter sets of $\{\alpha_{\rm E},\sigma_{\rm s}\}$ (magenta stars in Figure \ref{fig:contours}). The shaded region marks the bandpasses of the FAST and Arecibo telescopes.
\label{fig:sim_beta}}
\end{figure}

Adopting optimal parameter sets of $\{\alpha_{\rm E}, \sigma_{\rm s}\}$, Figure \ref{fig:sim_beta} displays the $\nu^{\rm sim}_{\rm p}$ and $\beta^{\rm sim}$ distributions of the simulated samples. The derived $\nu^{\rm sim}_{\rm p}$ distribution of the simulated FAST sample peaks at 0.87 GHz and 1.57 GHz, being consistent with the pre-set $\nu_{\rm p}$ distribution of the 1st and 2nd fringe components. The bandpass of the FAST telescope spans the two peaks. Most simulated FAST bursts ($75\%$) are of $\nu^{\rm sim}_{\rm p}$-off-band bursts. The inferred $\beta^{\rm sim}$ is dramatically different from burst to burst, ranging from -8 to 10. The $\nu^{\rm sim}_{\rm p}$ distribution of the simulated Arecibo sample, which covers from 1 GHz to 1.85 GHz, is broader than the $\nu^{\rm obs}_{\rm c}$ distribution. Most simulated Arecibo bursts ($65\%$) are $\nu^{\rm sim}_{\rm p}$-in-band burst. The $\beta^{\rm sim}$ distribution ranges from -8 to 12, but the $\beta^{\rm sim}$ values of about one-third of simulated Arecibo bursts are $\sim 0$, indicating that $\nu^{\rm sim}_{\rm p}$ of these bursts are very closed to the central frequency of the Arecibo telescope or the characterized frequency range $[\lambda_1, \lambda_2]$ is completely in the bandpass. The broad frequency range of the GBT telescope (4-8 GHz) makes the most simulated bursts $\nu^{\rm sim}_{\rm p}$-in-band bursts, and their $\beta^{\rm sim}$ values are $\sim 0$.

\section{Discussion}

\begin{figure}[!htbp]
\centering
\includegraphics[width=0.5\linewidth]{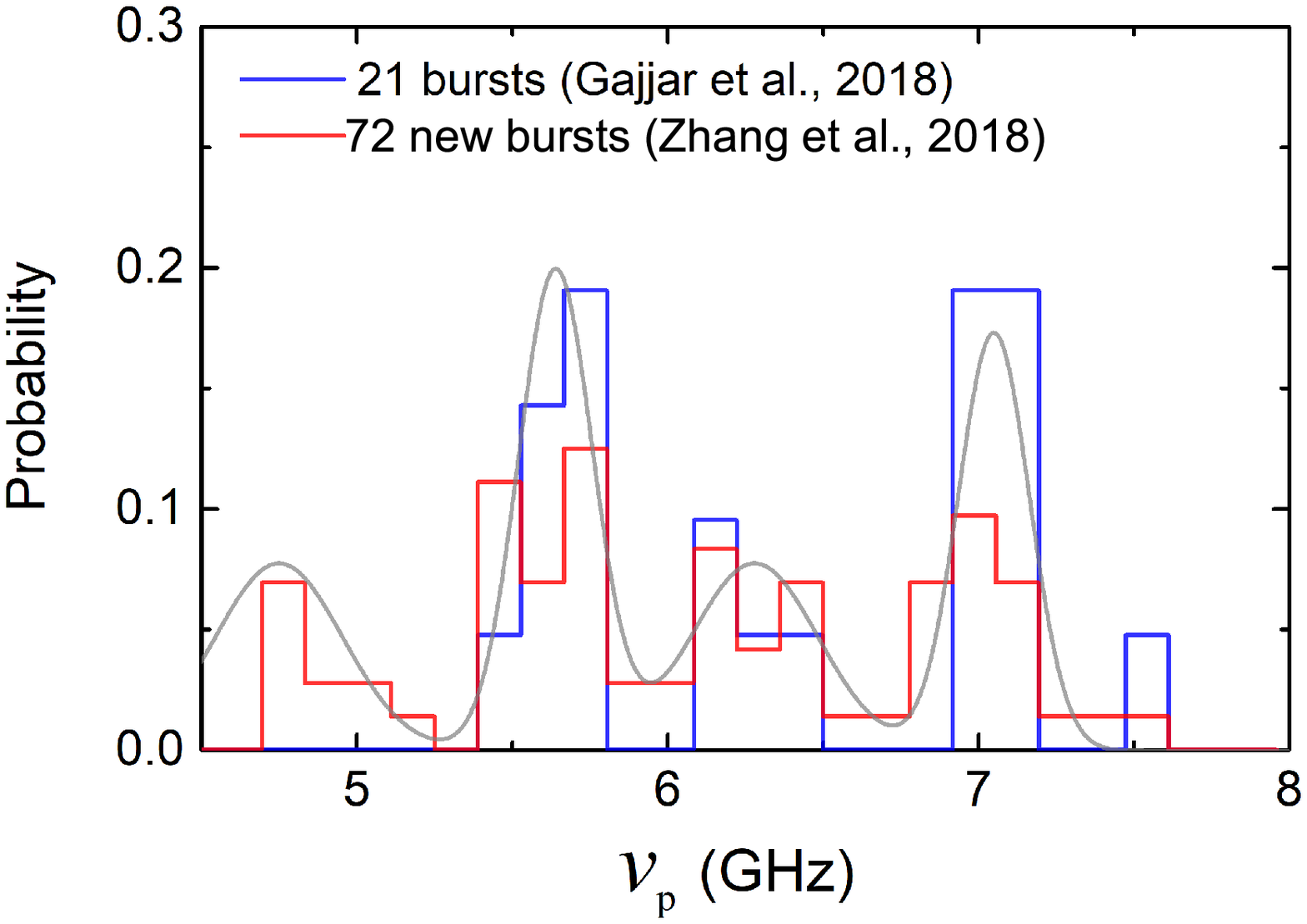}\hspace{-0.1in}
\includegraphics[width=0.5\linewidth]{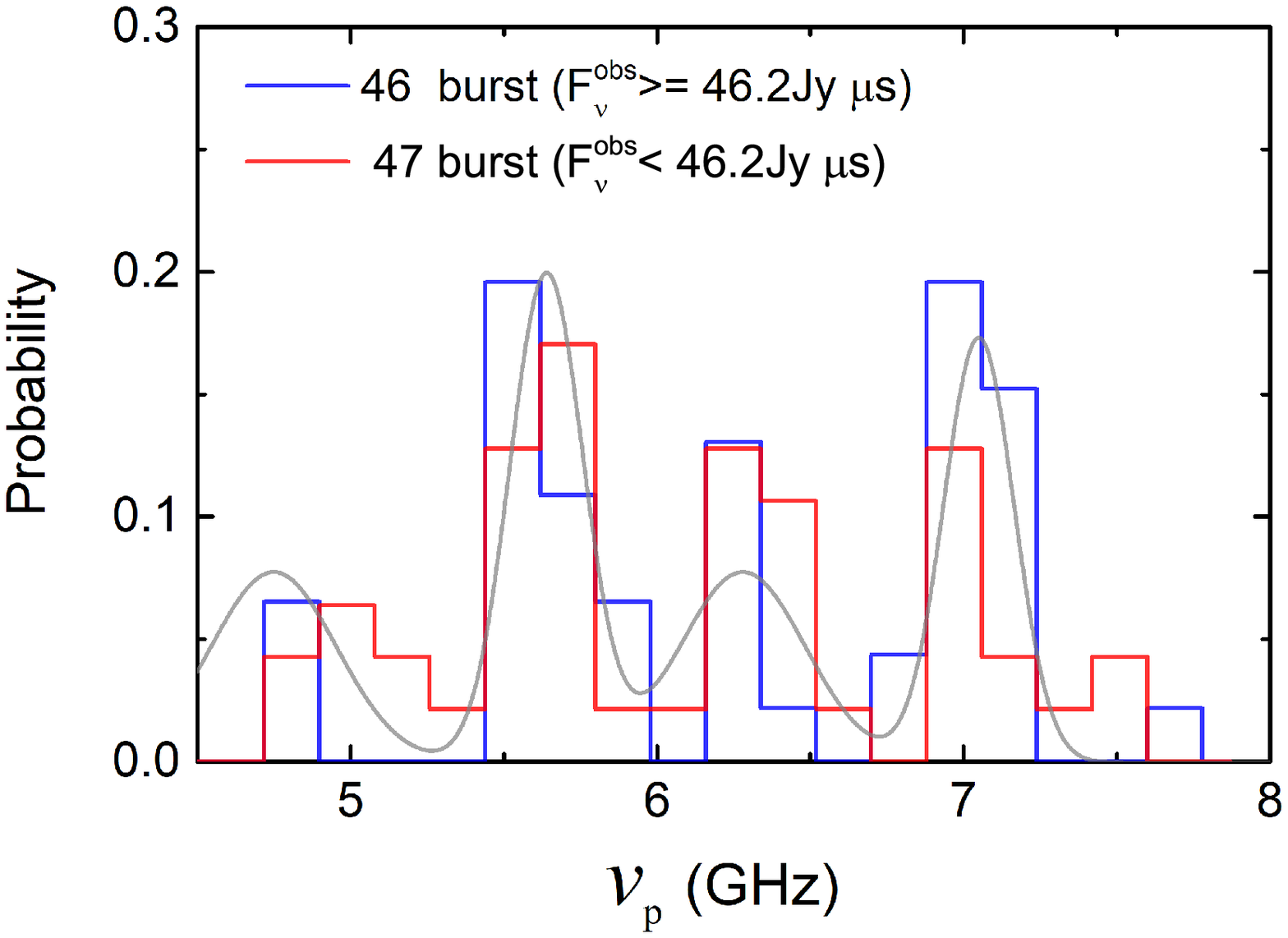}\vspace{-0.1in}

\caption{Comparisons of the $\nu_{\rm p}$ histograms between different sub-samples in the GBT sample:{\em left panel---} 21 bursts found by \cite{2018ApJ...863....2G} through a regular method {\em vs.} extra 72 bursts discovered by \cite{2018ApJ...866..149Z} through a neural network machine learning algorithm from the same GBT data; {\em right panel---} low- {\em vs.} high-fluence burst samples by averagely divided the total sample with a division $F_{\rm p}^{obs}$ $\geq$46.2 Jy $\mu$s. The gray line is the constructed $\nu_{\rm p}$ distribution for the total sample as shown in Figure \ref{fig:obs}.  
 \label{fig:test}}
\vspace{-0.2cm}
\end{figure}

\subsection{Putative Spectral Fringe Pattern and Narrow Spectral Width}
As shown in Figure \ref{fig:obs}, we construct a modulated $\nu_{\rm p}$ distribution for the bursts of FRB 20121102A based on the Arecibo and GBT observations in L-band and C-band. The interval among the $\nu_{\rm p}$-probability peaks is $\sim 0.8$ GHz. Such a discrete $\nu_{\rm p}$ distribution suggests that the radiations of FRB 20121102A in a broad frequency range show a fringe pattern. As mentioned in \S \ref{sec:intro}, a search for simultaneous bursts of FRB 20121102A in different frequencies indicates that the bursts are active in some preferred frequency bands, likely favoring the putative $\nu_{\rm p}$ fringe pattern derived from broadband (4-8 GHz) observations with the GBT telescope.

The putative $\nu_{\rm p}$ fringe pattern and the narrowness of the radiating spectrum should give insight into the radiation physics of FRBs. The proposed radiation models are classified into two groups, i.e. synchrotron maser in the relativistic shocks far away from the central engine (\citealp{2014MNRAS.442L...9L,2017ApJ...843L..26B,2019MNRAS.485.4091M} for the case of magnetized shocks and see \cite{2017ApJ...842...34W,2021ApJ...922...98D} for the case of weakly magnetized shocks) and the coherent curvature radiation (CR) or coherent inverse Compton scattering (ICS) of bunching electrons close-in the magnetosphere \citep{2018ApJ...868...31Y,2022ApJ...925...53Z}. 
Although the CR spectrum of a single bunch appears to oscillate at a typical narrow frequency and shows a discrete structure, the collective CR spectrum of bunches is characterized as several power-law segments in a broad frequency range \citep{2018ApJ...868...31Y}. The observed narrow spectra of FRBs are thought to be due to the absorption of low frequency radio emission, but the fringe pattern of $\nu_p$ cannot be expected. In the ICS radiation model, the characteristic frequency of the coherent emission depends on the Lorenz factor of the bunching electrons, as well as the frequency and the incident angle of the seed photons, which is possible to produce a narrow spectrum if all of these parameters take typical values in the framework the model \citep{2022ApJ...925...53Z}. Nonetheless, this model does not predict such a fringe pattern of $\nu_p$ as well. For the synchrotron maser emission models, analysis of particle-in-cell (PIC) simulations by \cite{2019MNRAS.485.3816P} shows that the synchrotron maser produces a narrow radiation spectrum peaking at $\rm{a~ few} \times \nu_e$ in the shock frame, where $\nu_e$ is the frequency of the pair plasma ahead of the shock. However, the observed peak frequency requires a Lorentz transformation from the shock frame to the observer, and the Lorentz factor of the shock depends on the burst energy, the burst time scale, and the property of the interacting medium. Therefore, it also does not predict the fringe pattern of $\nu_{\rm p}$ in the synchrotron maser model. The fringe pattern of $\nu_{\rm p}$ found in this work, if it is true, strongly challenges the two kinds of models mentioned above. New radiation models should be considered for generating such a fringe pattern intrinsically. They should be independent of the external environment and propagation effects.  We should note that potential analogs of the spectral fringe pattern are also found in the high frequency interpulse of Crab and the zebra patterns in solar radio spectra \citep{2013A&A...552A..90K,2016JPlPh..82c6302E}. They may give some hints for revealing the nature of the spectral fringe pattern.

Several statistical and observational caveats to the apparent fringe pattern should be addressed. First, it is unclear whether the fringe pattern results from  artificial effects, such as the signal-to-noise ratio (or fluence) threshold, the detection algorithm, the frequency ranges, and the burst bandwidths. Note that among the 93 bursts in the GBT sample for our analysis, 21 bursts were previously reported in \cite{2018ApJ...863....2G} and extra 72 bursts were discovered in \cite{2018ApJ...866..149Z} through the narrow bandwidth search by the use of a neural network machine learning algorithm. The neural network algorithm may have biases toward certain frequency ranges and burst bandwidths. As explained in \cite{2018ApJ...866..149Z}, their neural network machine learning algorithm modulates the spectra of their training data set. This might have caused a bias if the modulations did not cover the potential parameter space fully. We check whether the $\nu_p$ fringe pattern persists in the samples of the 21 bursts and the 72 bursts. As shown in Figure \ref{fig:test}, one can see that the $\nu_p$ fringe pattern exists and is consistent in the two sub-samples at $\nu_p>5.5$ GHz. The K-S test gives a probability of $p_{\rm KS}$=0.23. The bursts with $\nu_p<5.5$ GHz are only found with the neural network algorithm. Second, it is uncertain whether the fringe pattern is suffered a bias of fluence threshold selection effect. Threshold of signal-to-noise ratio for burst searching is taken as 6 in both \cite{2018ApJ...863....2G} and \cite{2018ApJ...866..149Z}. We averagely separate the global GBT sample into high-fluence and low-fluence groups with a fluence division of $F$=46.2 Jy $\mu$s and compare the $\nu_p$ fringe pattern in the two groups. As shown in Figure \ref{fig:test}, the $\nu_p$ fringe pattern in the low- and high-fluence groups are also statistically consistent ($p_{\rm KS}$=0.39), indicating that the $\nu_p$ fringe pattern persists in both the low and high fluence bursts. Finally, the $\nu_p$ fringe pattern suffered a great risk of statistical fluctuation effect based on a small sample of bursts identified from the GBT data. Especially, it is quite uncertain to extrapolate the fringe pattern in the GBT band to low frequency ranges. Notice that 9 bursts detected with VLA in the frequency coverage of 2.5-3.5 GHz likely imply a $\nu_{\rm p}$ fringe in this frequency range \citep{2017ApJ...850...76L}. The spectra of four out of the 9 bursts (bursts 57623, 57643, 57645, and 57648) peak at $\sim 2.8$ GHz, being likely inconsistent with the inferred fringe pattern, which is a trough around 2.8 GHz as shown in the top left panel of Figure \ref{fig:obs}. The such small number of statistics still makes concerns about the apparent fringe pattern. In addition, the apparent fringe pattern is derived from individual bursts at different times observed with GBT, but not from the simultaneous bursts observed in a broad frequency band. Broadband simultaneous observations in dense frequency coverage with the Square Kilometre Array (SKA), which has an extreme sensitivity and broad bandpass ($50\,\mathrm{MHz}\sim 20\,\mathrm{GHz}$, see the document available at the SKA website \footnote{ https://www.skatelescope.org/}; see also \citealt{2009IEEEP..97.1482D}) should offer an opportunity to verify this fringe pattern.

From Figure \ref{fig:contours}, the preferred $\sigma_{\rm s}$ value constrained with the FAST sample is $\sigma_{\rm s}=0.18^{+0.28}_{-0.06}$ GHz, and $\sigma_{\rm s} \leq 500$ MHz (the best $\sigma_{\rm s}$=0.17 GHz) with the Arecibo sample.  As shown in Figure \ref{fig:obs}, the spectral widths of the bursts in the FAST and Arecibo samples are $\Delta\nu^{\rm FAST}=0.1\sim 0.5$ GHz with a typical value of $\Delta\nu=0.22$ GHz and $\Delta\nu^{\rm Arecibo}=0.1\sim 0.7$ GHz with a typical value of 0.31 GHz. 
Taking $2\sigma_{\rm s}$ as the spectral width of the simulated bursts, we find that it is comparable to the observed ones. Note that the interval of the $\nu_{\rm p}$ fringes is $\sim 0.8$ GHz, which is longer than the derived spectral width. This is reasonable since the observed $\nu_{\rm p}$ fringes indicate that the radiating spectrum should be narrower than the fringe interval, otherwise the fringes could be smeared out by the broad radiating spectrum. The narrow spectrum feature makes the observed spectral shapes among bursts dramatically different, highly depending on the $\nu_{\rm p}$ is in or out of the bandpass. The diverse and time-varied spectral indices among bursts observed with narrow-bandpass telescopes (such as the FAST and Arecibo telescopes; \citealp{2016Natur.531..202S}) are well explained with the narrow spectral width and the fringe pattern of $\nu_{\rm p}$. 

\subsection{Intrinsic E-function of the FRB 20121102A bursts}
\begin{figure}[!htbp]
\centering
\includegraphics[width=0.5\linewidth]{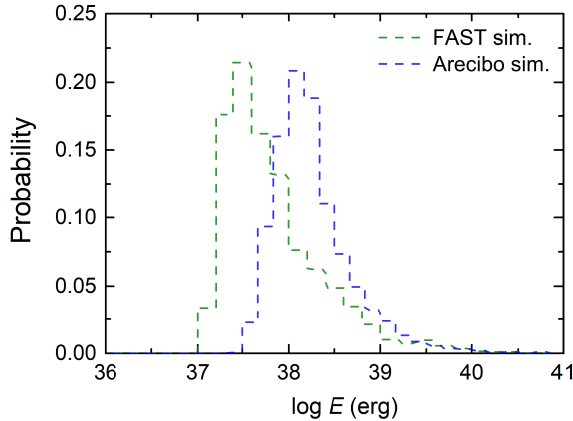}\hspace{-0.1in}
\caption{Distributions of the intrinsic energy $E$ of the simulated FAST and Arecibo samples.\label{fig:Eint}}
\vspace{-0.2cm}
\end{figure}

The observed E-distribution of FRB 20121102A is variable in different observational epochs \citep{2016ApJ...833..177S,2016Natur.531..202S,2017Natur.541...58C,2017ApJ...846...80S,2018Natur.553..182M,2018ApJ...866..149Z,2019ApJ...877L..19G}. \cite{2021Natur.598..267L} reported a two-component distribution of the specific energy distribution at $\mu_{\rm c}$ for the FAST sample, but it does not show up in the Arecibo and GBT samples, as shown in Figure \ref{fig:obs}. Our simulation analysis shows that the observed $E-$distribution of the three samples can be well reproduced by modeling the intrinsic E-function with a single power-law function. We compare the burst energy distributions of the simulated FAST and Arecibo samples \footnote{We do not add in the intrinsic E-distribution of the simulated GBT sample for comparison since the $\alpha_{\rm E}$ and $\sigma_{\rm s}$ values lose constraints with the GBT sample.} in Figure \ref{fig:Eint}. With a higher sensitivity, the intrinsic burst energy of the simulated FAST sample is lower than that of the simulated Arecibo sample, but no bimodal feature is observed.

Note that the specific energy of a burst depends on the specific frequency. For a narrow spectrum with a Gaussian profile, its $E_{\nu_{\rm p}}$ would be a reasonable representation of the burst energy. However, the bandpass of the FAST telescope covers the valley between the fringe peaks at $\nu_{\rm p}=0.87$ and $\nu_{\rm p}=1.57$ GHz. The $\nu_{\rm p}$ for most bursts in the simulated FAST sample is out of the FAST bandpass. Therefore, the $E_{\nu_{\rm c}}$ value is not representative of the intrinsic burst energy. Our simulations show that FAST can detect these $\nu_{\rm p}$-out-band bursts with its high sensitivity (see Figure \ref{fig:sim_beta}). This should make an excess of the low energy bursts. Therefore, the bimodal $E_{\nu_{\rm c}}$ distribution of the FAST sample should result from the intrinsic $\nu_{\rm p}$ fringe pattern and the narrow radiating spectrum as well as the detection of $\nu_{\rm p}$-out-band bursts with a high fluence sensitivity of the FAST telescope. This is also supported by the fact that the bimodal feature disappeared by calculating the burst energy over the bandpass \citep{2021ApJ...920L..18A}.

\section{Conclusions}
In this paper, we have investigated the intrinsic radiating spectrum and energy distribution of FRB 20121102A through Monte Carlo simulations by adopting multi-frequency observations with the FAST (1.05-1.45 GHz), Arecibo (1.15-1.73 GHz), and GBT (4-8 GHz) telescopes. Our results are summarized below.
\begin{itemize}
    \item Using the GBT sample of FRB 20121102A, we find a fringe feature of the $\nu^{\rm obs}_{\rm p}$ distribution, which can be fitted with a series of normal functions with peaks at 4.75, 5.58, 6.28, 7.06 GHz, indicating that the bursts are active in these preferred frequencies. The intervals among the peaks are $\sim 0.8$ GHz. Combing the GBT and Arecibo samples, we construct the intrinsic $\nu_{\rm p}$ fringe pattern in 0.5-8 GHz. Current simultaneous broadband observations in sparse frequency coverage still do not reveal simultaneous burst activities in these preferred $\nu_{\rm p}$.
    \item We investigate the intrinsic E-distribution and radiating spectrum profile through Monte Carlo simulations. By modeling the intrinsic energy distribution as a single power-law function and depicting the spectrum profile with a Gaussian, our simulations show that the maximum likelihood parameter set $\{\alpha_{\rm E},\sigma_{\rm s}\}$ derived from the FAST and Arecibo samples are $\{1.82,0.18\}$ and $\{2.09,0.17\}$, respectively. The $\sigma_{\rm s}$ value well agrees with the observed spectral range of the FAST and Arecibo observations. The observed E-distributions of the three samples are well reproduced with the parameter sets. Especially, the bimodal E-distribution and the significant spectral slope variation of the FAST sample result from the spectral fringe pattern and narrow Gaussian spectral profile as well as the detection of $\nu_{\rm p}$-out-band bursts with high sensitivity of the FAST telescope.
\end{itemize}
In conclusion, our results suggest that the $\nu_{\rm p}$ among bursts of FRB 20121102A illustrates a fringe pattern in a broad energy frequency range, and the variations of the observed E-distribution and  spectral slope are physically due to both the intrinsic $\nu_{\rm p}$ fringe pattern and the narrowness of the radiating spectrum, and observationally due to the bandpass selection and sensitivity of different telescopes.


\section*{acknowledgments}
We very much appreciate thoughtful and constructive comments and suggestions from the referee. We also thank Di Li, Bing Zhang, Xue-Feng Wu, Fa-Yin Wang, Pei Wang, Wei-Yang Wang, and Yuan-Pei Yang for their helpful discussion. We acknowledge the use of the public data from the FAST/FRB Key Project. F.L. is supported by Shanghai Post-doctoral Excellence Program. E.W.L and J.G.C. are supported by the National Natural Science Foundation of China (grant Nos.~12133003). C.M.D. is supported by the National Natural Science Foundation of China (grant No. 12203013) and the Guangxi Science Foundation (grant Nos. 2021AC19263).

%





\end{document}